%% file: TOSEM_2025_PredictingSoftwareChanges.tex
\PassOptionsToPackage{table,dvipsnames}{xcolor}
\documentclass[manuscript,screen,nonacm]{acmart}
\AtBeginDocument{%
  }

\usepackage{minted}
\usepackage{multirow}
\usepackage{tcolorbox}
\usepackage{pifont}
\usepackage{mdframed}
\usepackage{adjustbox}
\usepackage{listings}
\usepackage{subcaption}
\usepackage{float}
\usepackage[nameinlink]{cleveref}
\usepackage{siunitx}
\usepackage[inline]{enumitem}

\newboolean{showcomments}
\setboolean{showcomments}{true}
\ifdefined\notodocomments
  \setboolean{showcomments}{false}
\fi
\ifthenelse{\boolean{showcomments}}
 { \newcommand{\mynote}[2]{
      \fbox{\bfseries\sffamily\scriptsize#1}
        {\small$\blacktriangleright$\textsf{\emph{#2}}$\blacktriangleleft$}}}
        { \newcommand{\mynote}[2]{}}

\newcommand{\wcircle}[1]{\ding{\numexpr171 + #1}}

\newcommand{\highlight}[1]{%
\begin{tcolorbox}[leftrule=1mm,rightrule=1mm,toprule=0mm,bottomrule=0mm,left=0pt,right=0pt,top=0pt,bottom=0pt, colback=gray!30, colframe=gray!90]
#1
\end{tcolorbox}%
}

\newcommand{\gpt}{GPT4.1-mini}

\definecolor{delim}{RGB}{20,105,176}
\definecolor{numb}{RGB}{106, 109, 32}
\definecolor{string}{rgb}{0.64,0.08,0.08}

\lstdefinelanguage{json}{
    numbers=left,
    numberstyle=\small,
    rulecolor=\color{black},
    showspaces=false,
    showtabs=false,
    breaklines=true,
    postbreak=\raisebox{0ex}[0ex][0ex]{\ensuremath{\color{gray}\hookrightarrow\space}},
    breakatwhitespace=true,
    basicstyle=\ttfamily\small,
    upquote=true,
    morestring=[b]",
    stringstyle=\color{string},
    columns=fullflexible,
    literate=
     *{0}{{{\color{numb}0}}}{1}
      {1}{{{\color{numb}1}}}{1}
      {2}{{{\color{numb}2}}}{1}
      {3}{{{\color{numb}3}}}{1}
      {4}{{{\color{numb}4}}}{1}
      {5}{{{\color{numb}5}}}{1}
      {6}{{{\color{numb}6}}}{1}
      {7}{{{\color{numb}7}}}{1}
      {8}{{{\color{numb}8}}}{1}
      {9}{{{\color{numb}9}}}{1}
      {\{}{{{\color{delim}{\{}}}}{1}
      {\}}{{{\color{delim}{\}}}}}{1}
      {[}{{{\color{delim}{[}}}}{1}
      {]}{{{\color{delim}{]}}}}{1},
}

\begin{document}

\title[Neural Change Prediction]{Neural Change Prediction: Relating~Software~Changes~to~Their~Effects~and~Vice~Versa}

\author{Laura Plein}
\email{laura.plein@cispa.de}
\orcid{0009-0000-9424-6762}
\affiliation{%
  \institution{CISPA Helmholtz Center for Information Security}
  \city{Saarbrücken}
  \country{Germany}
}

\author{Souhila Zidane}
\email{sozi00002@stud.uni-saarland.de}
\orcid{0009-0008-4256-3051}
\affiliation{%
  \institution{Saarland University}
  \city{Saarbrücken}
  \country{Germany}
}

\author{Jordan Samhi}
\authornote{Most of this work was conducted while the author was a postdoctoral researcher at CISPA.}
\email{jordan.samhi@uni.lu}
\orcid{0000-0001-6052-6184}
\affiliation{%
  \institution{University of Luxembourg}
  \city{Luxembourg}
  \country{Luxembourg}
}

\author{Andreas Zeller}
\email{andreas.zeller@cispa.de}
\orcid{0000-0003-4719-8803}
\affiliation{%
  \institution{CISPA Helmholtz Center for Information Security}
  \city{Saarbrücken}
  \country{Germany}
}

\definecolor{framecolor}{RGB}{0,0,0}
\definecolor{backcolor}{RGB}{245,245,245}
\definecolor{titlecolor}{RGB}{70,70,70}

\mdfdefinestyle{niceframe}{
    linewidth=1pt,
    linecolor=framecolor,
    backgroundcolor=backcolor,
    roundcorner=5pt,
    innertopmargin=\baselineskip,
    frametitlefont=\bfseries\sffamily\color{titlecolor},
    frametitlealignment=\raggedright,
    frametitlebackgroundcolor=backcolor,
    shadow=true,
    shadowsize=3pt,
    shadowcolor=black!20,
}

\renewcommand{\shortauthors}{Plein et al.}

\begin{abstract}

Much of software development revolves around understanding the relationship between \emph{software changes} and their \emph{effects}.
\emph{Software debugging,} for instance, is about finding a change to
software such that a given effect (the failure) no longer occurs.
Similarly, \emph{software evolution} focuses on introducing changes to achieve a desired effect.
If we could \emph{learn} and \emph{predict} the effects of software changes (and conversely learn and predict which software changes would produce a particular effect)
such predictions could benefit several areas of software engineering.
But while recent advances in artificial intelligence have shown great promise in software engineering tasks, interpreting and predicting the semantics of code \emph{without actually executing it} remains a big challenge for machine learning models.

In this paper, we present \emph{Neural Change Prediction}, a novel and fundamental technique to \emph{learn} and \emph{predict} the associations between software changes and their \emph{dynamic} effects on program behavior.
Specifically, for a given program and a set of test inputs, we automatically apply numerous \emph{mutations} to the source code and configuration files of a program and then observe how these changes alter the program's \emph{output}.
From these (changes to software, changes in behavior)-pairs, we then create models that
\begin{enumerate}
\item for a desired change in behavior, predict \emph{where} and \emph{how} the code should be changed (feature localization, software evolution, and software repair); and
\item for a given code change, predict how this code change \emph{affects the output} (effect prediction).
\end{enumerate}

We have conducted a detailed case study on CSS configuration files and a large-scale evaluation on Python programs to demonstrate the generality and wide applicability of Neural Change Prediction.
Central findings include:
\begin{enumerate}
  \item Given a desired change in appearance (specified in natural language), Neural Change Prediction successfully predicted the correct changes to CSS configuration files with an accuracy of up to~95\% (fine-tuned GPT-4.1, general learning); project-specific learning boosts this to 100\% accuracy.
  \item Given a desired change in behavior of Python (specified as a change in output for a given input), Neural Change Prediction successfully predicted the correct \emph{change location} with an accuracy of 82.6\% (fine-tuned GPT-4.1, single mutation) and the \emph{exact change} with an accuracy of 71.6\% (fine-tuned GPT-4.1, single mutation).
  \item Given a code change, Neural Change Prediction successfully predicted the change in output with an accuracy of 95\% (fine-tuned GPT-4.1, single mutation); predicting the effect of \emph{multiple} mutations has an accuracy of 99\%.
\end{enumerate}
To put these numbers into context, current LLM systems struggle with all of these tasks, achieving only 10\%~to~33\% accuracy.
All the predictions can be trivially validated by running the (changed) program, mitigating the risk of false predictions.

While Neural Change Prediction requires numerous mutations (and thus numerous executions of the program under test), Neural Change Prediction is fully automatic and does not require any prior knowledge of the code or its semantics, making it applicable to any software artifact that can be executed and whose output can be observed.
\end{abstract}

\ccsdesc[500]{Software and its engineering~Software defect analysis}
\ccsdesc[500]{Software and its engineering~Software testing and debugging}
\ccsdesc[100]{Software and its engineering~Maintaining software}
\ccsdesc[100]{Software and its engineering~Automatic programming}

\keywords{Software Changes, Code Generation, Change Prediction}

\maketitle

\input{sections/intro}

\input{sections/approach}

\input{sections/caseStudies}
\input{sections/experiments}

\input{sections/discussion}

\input{sections/rw}

\input{sections/conclusion}

\begin{acks}
This work is funded by the European Union (ERC S3, 101093186). Views and opinions expressed are however those of the author(s) only and do not necessarily reflect those of the European Union or the European Research Council. Neither the European Union nor the granting authority can be held responsible for them.
\end{acks}

\bibliographystyle{ACM-Reference-Format}
\bibliography{references}

\end{document}

%% file: sections/intro.tex
\section{Introduction}

How can I change software to achieve a particular effect?
And what will be the effect of this particular software change?
Relating \emph{software changes} and their \emph{effects} is one of the central challenges of software development.
In \emph{software debugging,} we need to find a change that gets rid of an unwanted effect, e.g., the fix addressing a program failure.
When \emph{evolving software,} we want to ensure that the changes we made have the desired effect (say, a new feature), again without introducing side effects.
Both debugging and evolution typically start with \emph{fault localization} or \emph{feature localization}---that is, finding the code that causes a particular effect; and ``causes'' again means that if we alter said code, the effect will be changed.
Understanding the effects of software changes, and which changes are needed to achieve a particular effect, thus are key to efficient and trustable software maintenance.

In the past decades, a number of automated techniques have been developed that assist developers in relating software changes and their effects.
\emph{Change impact analysis}~\cite{arnold1996software} determines which parts of a program could possibly be affected by a change.
\emph{Fault localization}~\cite{wong2016survey} explores the code locations whose execution correlates with failure, making them candidates for fixing the bug.
\emph{Automated program repair}~\cite{le2019automated} searches for possible \emph{code changes} to get rid of a failure.
\emph{Feature localization}~\cite{dit2013feature} determines code locations related to a particular effect---that is, locations that can be \emph{changed} to alter said effect.
While all these techniques have been shown to help developers, all of them have still much room for improvement.

In this paper, we present a novel and \emph{fundamental} technique to \emph{learn} and \emph{predict} the associations of software changes and effects.
Specifically, for a given program and a set of inputs, we automatically apply numerous \emph{synthetic changes} to the program code and observe the effects of these mutations on the program output.
What we thus obtain is an arbitrary large set of pairs of code changes together with output changes that we can train a \emph{learner} from.
Thus trained, our learner can then, given a change to the code, predict its change to the output \emph{without executing the program.}
Better yet, we can train the learner in the inverse direction---with (change to the output, change to the code) pairs---and then, given a desired change to the output, \emph{predict a change to the code} that should result in the desired change in behavior.

Applying thousands of mutations to code or configuration files and observing their effects is expensive in terms of compute resources, but also fully automated.
In contrast to traditional techniques, though, Neural Change Prediction needs no prior knowledge of the code or its semantics---neither as part of a static or dynamic analysis, nor as part of a machine learning model.
Instead, it is set to \emph{learn the semantics from scratch,} by pure observation and making associations.
This makes Neural Change Prediction \emph{language-agnostic} in the sense that it can be applied to any software artifact that can be executed and whose output can be observed.

Let us illustrate Neural Change Prediction with a simple example.
\emph{Spurgeon}\footnote{\url{https://themewagon.com/themes/spurgeon/}} is a theme for websites, consisting of HTML, CSS, and JavaScript code to be used as templates.
The 1.0.0 template is defined by 11,677~lines of code: 5,110~lines of HTML, 6,134~lines of CSS3 code, and 433~lines of JavaScript.
Changing any aspect of the template, such as the color of any element, the size of any font, or the layout of any page, is a matter of making the right change---somewhere in these 11,677 lines of code.\footnote{Neural Change Prediction works on code and configuration files alike. While CSS and HTML are not programming languages in the strict sense, they are not part of the input, but part of code whose effects can be observed.}

Using Neural Change Prediction, we would train a model that can predict how to change the Spurgeon template code to achieve a desired effect.
For this training, we design a set of \emph{mutation operators} to the code of the Spurgeon template.
Such mutations might change the color of a text element from \texttt{red} to \texttt{blue} by replacing the corresponding CSS value; they also might change the size of a font by replacing the corresponding CSS rule \texttt{font-size:12px} with \texttt{font-size:14px}.
For each such mutation, we record its \emph{intent} in natural language (e.g., ``change color from red to blue''), the original code, and the changed code.
We repeat this process for thousands of random mutations on the Spurgeon template, and thus obtain a large set of triples (original code, intent, mutated code).
We use this data to fine-tune a \emph{language model} that can predict the code change that is most likely to achieve a desired effect.
To this end, we use templates of the form shown in \Cref{lst:spurgeon_template}, where we use the original code and the intent as input, and the mutated code as output.

\begin{lstlisting}[language=json,caption=Example of a training example for the Spurgeon template,label=lst:spurgeon_template,float=b,captionpos=b,showspaces=false,showstringspaces=false]
{
  "messages": [
    {
      "role": "system",
      "content": "Be a helpful web developer and make the right CSS changes."
    },
    {
      "role": "user",
      "content": "I have the following HTML element: \n\n 'element_html', and its corresponding CSS code: \n\n 'original_css'. \n\n Please make the corresponding CSS change: 'user_intent'"
    },
    {
      "role": "assistant",
      "content": "Here is the corresponding CSS code:\n\n 'mutated_css'"
    }
  ]
}\end{lstlisting}

Once trained on the Spurgeon triplets, our model can then be used to \emph{predict} the code change that is most likely to achieve a desired effect.

\begin{figure}[h]
    \centering
    \includegraphics[width=\textwidth]{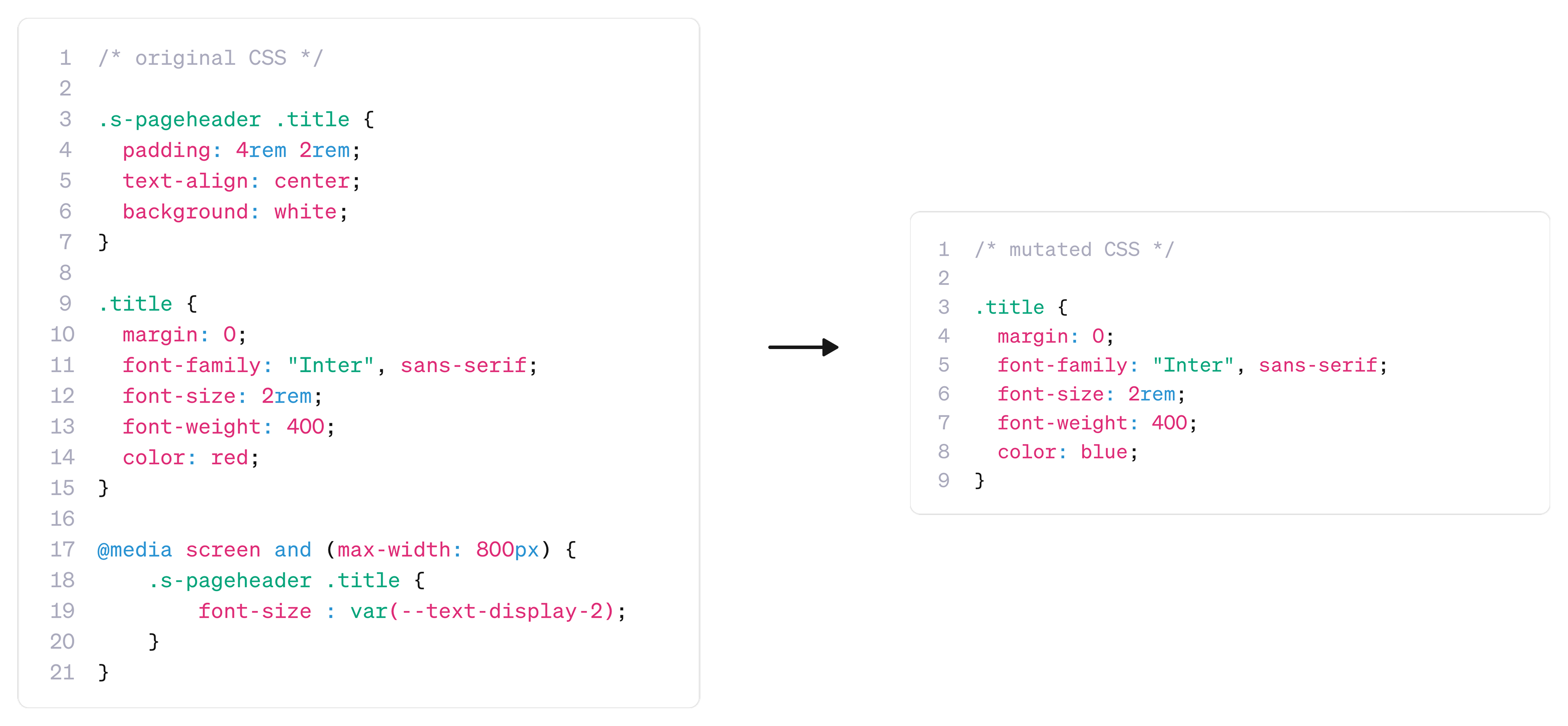}
    \caption{Selection and mutation of the relevant CSS rule from candidate CSS rules.}
    \label{fig:spurgeon-example}
\end{figure}

For instance, if users want to change the color of the title from red to blue, they would
\begin{enumerate}
  \item Select the title element in a rendered page;
  \item Describe the desired change in natural language, e.g., ``please change color from red to blue''; and
  \item Ask the model to predict the corresponding code change.
\end{enumerate}
From the selected rendered output element (the title), one can easily determine the CSS and HTML code that determines the title.
Such ``original code'' will still consist of several individual rules (especially in CSS) that determine the features of the title; and thus it is not obvious which of these rules needs to be changed to achieve the desired effect.
This original code, together with the intent (``please change color from red to blue''), is then fed into the model, which predicts the mutated code that is most likely to achieve the desired effect. 
\Cref{fig:spurgeon-example} illustrates this process with a simplified CSS example.

In our experiments with CSS templates, we show that a trained model can predict the correct code change with an accuracy of up to 95\% (fine-tuned \gpt, general learning across all templates).
Project-specific learning (from one template only, such as Spurgeon) boosts this to 100\% accuracy.

Neural Change Prediction is not limited to CSS or HTML code, though---any setting that can be expressed as (original code, intent, mutated code) triplets can be used for training and prediction.
In particular, we also apply Neural Change Prediction to Python code, where the original code and the mutated code are Python code snippets, and the intent is a change in output for a given input; here as well, Neural Change Prediction successfully predicts change locations and exact changes when given a desired change in output.

Also, it is important to note that, all we need is
\begin{enumerate*}[label=(\arabic*)]
\item a set of mutation operators that mostly keep the code valid; and
\item a means to automatically execute the program and determine output changes.
\end{enumerate*}
The above process can thus be applied to any artifact that determines program execution---be it program code, configuration files, or settings.

In summary, we make the following contributions:
\begin{description}
    \item[Neural Change Prediction: A learning-based approach to associate software changes and effects.] We introduce Neural Change Prediction, the first approach to systematically apply synthetic changes~$C$ to program to assess and learn the impact of these mutations on the output~$O$.
    \item[Learning models for assessing the effect of a software change.] Neural Change Prediction learns a \emph{model} that given a software change~$C$ and a current output $O$, can predict how $C$~changes the output from $O$~to~$O'$.
    \item[Learning models for predicting software changes that produce a desired effect.] Given a current output~$O$ and a desired output~$O'$, a trained Neural Change Prediction model can predict the software change that changes the output from $O$~to~$O'$.
    \item[Assessing and predicting software changes beyond code.]
    Neural Change Prediction is \emph{language-agnostic} in the sense that no code analysis is required (only language-specific mutation operators) and that the approach can just as well be applied to configuration files or other mostly static inputs.
\end{description}
We see our work as a novel alternative to the traditional approaches of change impact analysis, fault localization, automated program repair, and feature localization---in some ways more general, in some ways more effective, but also at the cost of more compute resources.
The present work thus opens up a new avenue of research to learn and predict software changes and their effects.

The remainder of this paper is organized as follows: \Cref{sec:approach} presents Neural Change Prediction, our novel approach to learn the effect of software changes and the workflow from a user's perspective. \Cref{sec:case_studies} presents a case study on predicting website configurations with CSS which highlight the generality and wide application range of Neural Change Prediction. We introduce the research questions that will be evaluated in \Cref{sec:experimental_evaluation}. \Cref{sec:experimental_evaluation} further enumerates the curated datasets, and the evaluation metrics before reporting and discussing the results of the research questions. Threats to validity, current limitations, and opportunities for future work are elaborated in \Cref{sec:discussion}. \Cref{sec:rw} presents some related work. Finally, \Cref{sec:conclusion} concludes our work.

Neural Change Prediction and all of our experimental data are available on request.\footnote{We would like to avoid contamination of AI models through publicly available data. However, we are happy to make the data available to any human researcher.}

%% file: sections/approach.tex
 \section{Approach}\label{sec:approach}

In this section, we introduce a novel approach that trains project-specific models able to capture the complex relationship between code changes and behavior changes. 
Our goal is to develop a general, language-agnostic method that can: 
(1)~predict where a change is needed to achieve a desired behavioral change (\textbf{localization} prediction);
(2)~synthesize the necessary code change (\textbf{code change} generation); and 
(3)~predict the effect of a given code change (\textbf{effect} prediction).
To achieve this, we train models using features (i.e., \emph{signals}) collected at runtime from programs, thereby obtaining project-specific models that learn how synthesized changes to code (mutations) lead to changes in behavior (notably, the program output).

\begin{figure}[h!]
  \centering
  \begin{adjustbox}{width=.95\linewidth}
  \includegraphics[width=\linewidth]{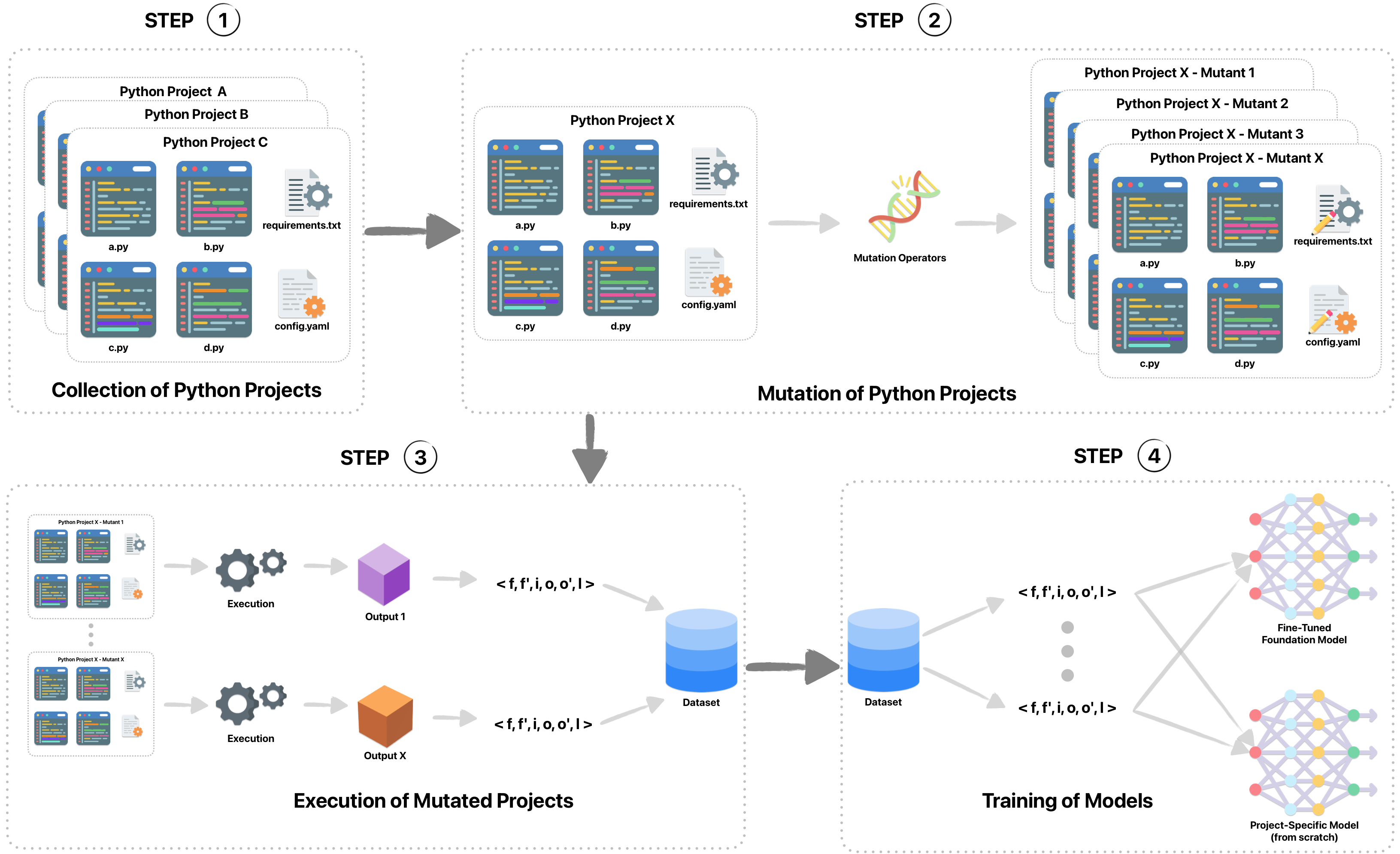}
  \end{adjustbox}
  \caption{Overview of Neural Change Prediction, illustrated with Python projects.  The pipeline is composed of four steps: (1)~collection of projects,  (2)~mutation of code and configurations, (3)~execution and vector representation $\langle f, f', i, o, o', l \rangle$, and (4)~training of project-specific and fine-tuned models. Here, $f$ denotes the original project,  $f'$ the mutated project,  $i$ the program input,  $o$ the output of the original project on input $i$,  $o'$ the output of the mutated project on the same input,  and $l$ the location of the applied mutation.} 
  \label{fig:approach_overview}
\end{figure}

\Cref{fig:approach_overview} depicts the overview of Neural Change Prediction, illustrated with an example on Python projects.
Our methodology is divided into four main steps:

\begin{enumerate}
  \item \textbf{Collection of Software Projects.}
    We begin by collecting a set of publicly available software projects. 
    In the example presented in \Cref{fig:approach_overview}, we focus on Python projects as those will be the target of our experimental evaluation in \Cref{sec:experimental_evaluation}. The additional projects collected for the case study will be introduced in \Cref{sec:cs_css}.

  \item \textbf{Dataset Augmentation via Mutations.}
    To augment our dataset and enable the training of project-specific models, 
    we automatically apply mutations to the collected code bases in order to generate additional project variants. A dedicated mutation tool for css has been implemented for this study as discussed in \Cref{sec:cs_css}. To generate and execute the Python mutants we used PyMut4SE\footnote{\url{https://github.com/LaPlei96/PyMut4SE}} 

  \item \textbf{Execution and Data Collection.}
    For each (original and mutated) project in our dataset, 
    we execute the program, in a dedicated execution environment, on a set of test inputs and collect the resulting outputs.
    Each execution is stored in the database with additional information about the mutation degree, mutation operator as well as error messages if the mutants' execution was not successful.
    
    From the collected mutants and execution features we can extract the information relevant for training, which in the context of a Python project can be in the form of a feature vector: $\langle f, f', i, o, o', l \rangle$
    where $f$ is the original project, $f'$ the mutated project, $i$ the input, 
    $o$ the output of the original project, $o'$ the output of the mutated project, 
    and $l$ the location of the mutation in the code. The remaining information will be leveraged to analyse and discuss the experimental results as well as investigate how specific factors such as the mutation degree impact the models performance.

  \item \textbf{Model training.}
  Using datasets constructed from our database, we train two types of models:
  \begin{enumerate}[label=(\roman*)]
  \item \textit{Open source models} are used in this work to demonstrate the feasibility and applicability of Neural Change Prediction. 
  \item \textit{Commercial models} fine-tuned with the same behavioral information highlight the full potential of this work by achieving great performance across domains.

  \end{enumerate}
\end{enumerate}

\subsection{Choice of Models}

Several model architectures can be leveraged to \emph{relate code changes to behavior changes}. 
Recent years have witnessed an explosion of Machine Learning techniques in software engineering~\cite{austin2021program,xia2023automated,plein2024automatic}, 
with Large Language Models (LLMs) becoming increasingly prevalent. Most of these approaches, however, operate primarily on program code as text, and predict token sequences in specific contexts, while the dynamic semantics of code remains far less explored.
Recently, multimodal models~\cite{jiang2025viscodex,wang2025code} have gained significant attention because they can jointly leverage natural language and visual inputs. 
They are particularly relevant for studying the effects of software changes in visual contexts. For example, modifications to CSS can directly affect how a website is rendered, and these visual changes can be effectively captured through website screenshots.

In our study, we adopt a dual strategy. 
First, we \emph{fine-tune general foundation models}, 
taking advantage of their strong syntactical knowledge of code 
and the large-scale training they have already received on code and natural language. 
During fine-tuning, we inject Python-specific behavioral information 
derived from dynamic executions, which enables the models to go beyond syntax 
and capture the semantics of code changes in their concrete runtime contexts.
Second, in our case study, we further train \emph{project-specific models}. 
Because our methodology employs systematic mutations, 
we can generate an arbitrarily large number of (change--effect) pairs, 
and ensure that we can gather a representative amount of mutants to collect sufficient data to learn the dynamic behavior of a project. Thus, the initial data scarcity of the project does not limit training. Project-specific models could be trained in the future for arbitrary project, given that a mutation tool in the projects programming language is available.

We consider commercial and open-source models from different families. While the research questions are investigated by fine-tuning models from OpenAI (\gpt\footnote{\url{https://openai.com/index/gpt-4-1/}}, across the experiments of our case study, we additionally consider GPT-oss\footnote{\url{https://openai.com/index/introducing-gpt-oss/}}), CodeLLama\footnote{\url{https://www.llama.com/}} from Meta, and Qwen\footnote{\url{https://qwen.ai/apiplatform}}.

Additionally, traditional machine learning algorithms provide a comparison baseline and allow us to study how simpler models perform when supplied with behavioral data. 
The encoding of data depends on the chosen architecture: 
while some models consume natural language prompts directly, 
others require tokenized representations or structured encodings of our feature vectors. 
Nevertheless, the underlying structure of our dataset remains consistent 
across all models. 
Further details on the encoding strategies and model configurations 
will be provided in \Cref{sec:case_studies} and \Cref{sec:experimental_evaluation}. 

\subsection{Note on Generality}

It is important to highlight that Neural Change Prediction is \emph{highly dependent on the dataset} used to train the models.
While we demonstrate it on Python projects in this work (and later on CSS in \Cref{sec:case_studies}), 
the methodology is not restricted to Python.
The core idea is to \emph{learn program behavior from signals observed at runtime}, 
independently of the programming language or framework.
Since the approach is generic, more details about the datasets used in our experiments 
will be provided in the subsequent sections, 
where we explain how we collected and mutated projects 
and how we trained our models.

\subsection{Workflow of Neural Change Prediction}

While \Cref{fig:approach_overview} illustrates the end-to-end process of building and training our models, 
\Cref{fig:workflow} focuses on how the trained models are used in practice. 
Given a program, an input, and the corresponding output, our model addresses three tasks 
that correspond to common software engineering challenges:

\begin{figure}[h]
  \centering
  \begin{adjustbox}{width=.95\linewidth}
  \includegraphics[width=\linewidth]{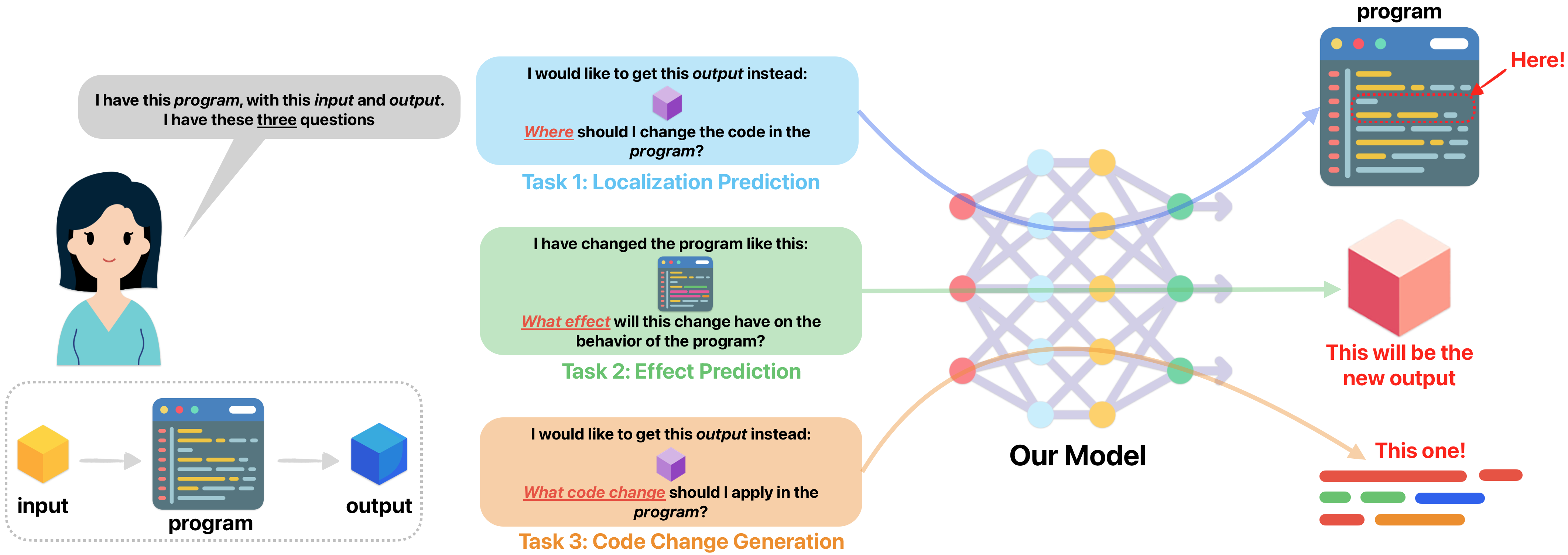}
  \end{adjustbox}
  \caption{Workflow of our model. 
Given a program, its input, and its output, the model answers three questions: 
\wcircle{1}~\textbf{Localization Prediction} (Where should the code be changed to obtain a desired new output?), 
\wcircle{2}~\textbf{Effect Prediction} (What effect will a given code change have on the program's behavior?), and 
\wcircle{3}~\textbf{Code Change Generation} (What code change should be applied to obtain a desired output?). 
}
  \label{fig:workflow}
\end{figure}

\begin{enumerate}
    \item \textbf{Localization Prediction.}
    Developers may want to obtain a different output for a given input. 
    In this case, the model predicts \emph{where} in the program the code should be changed 
    to produce the desired output. 
    This task directly supports \emph{fault localization}.

    \item \textbf{Effect Prediction.}
    When a developer proposes a change to the program, 
    the model predicts the \emph{effect} of this change on the program’s behavior. 
    This allows to assess the potential consequences (intended and unintended) of a modification 
    before it is applied. 
    This process would enable a form of \emph{change impact analysis}.

    \item \textbf{Code Change Generation.}
    Finally, if a developer specifies a desired output for a given input, 
    the model can propose \emph{what code change} should be applied 
    to achieve the new behavior. 
    This corresponds to \emph{automated program repair}.
\end{enumerate}

\noindent
\textbf{A Simple Example.}
Collecting a large amount of (code change--behavior change) pairs and training models that capture their effects 
are only part of the challenge. 
To be truly usable, our model must support simple and intuitive interactions, to allow users to express their intent in \emph{natural language}.

Consider, for example, a \texttt{gcd()} function that, given the input \texttt{(2,4)}, 
returns \texttt{``4''} instead of the expected \texttt{``2''}. 
A developer might query the model as illustrated in~\Cref{lst:gcd}.

\begin{lstlisting}[language=json,caption=A simple interaction example,label=lst:gcd,float=b,captionpos=b,showspaces=false,showstringspaces=false]
{
  "messages": [
    {
      "role": "user",
      "content": "Here is a Python function: {ORIGINAL FUNCTION f}. 
                  When given the input {INPUT i}, it produces the output {OUTPUT o}. 
                  However, I would like it to produce {OUTPUT o'}. 
                  Can you provide the corrected Python code?"
    }
  ]
}\end{lstlisting}

In this scenario, the model must determine 
\emph{where} in the function the change should be applied (\textbf{localization} prediction), 
\emph{what effect} the change will have (\textbf{effect} prediction), 
and \emph{how} to modify the code to produce the desired output (\textbf{code change} generation).

To demonstrate the applicability and workflow of Neural Change Prediction in a real-world scenario, we conduct a \textbf{case study} that illustrates how project-specific models trained with our methodology 
can support developers in understanding and predicting the impact of software changes,
is presented in \Cref{sec:case_studies}.

%% file: sections/caseStudies.tex
\section{Case Study}
\label{sec:case_studies}

Our case study targets web development, and focuses on \emph{Predicting Website Configuration Changes in CSS}. 
As websites often require precise adjustments in styling and layout, 
developers may struggle to identify which configuration changes produce the desired visual outcome. 
Here, we train a website configuration specific model tailored to CSS configurations 
to learn the impact of style modifications and assist in generating the required changes. 
This case study demonstrates the versatility of Neural Change Prediction beyond Python, 
showcasing its application to configuration files and web technologies.

\subsection*{Overview}\label{sec:cs_css}  

Modern websites are heavily driven by the combination of the DOM (Document Object Model) and CSS (Cascading Style Sheets), which together define the structure and presentation of webpages. While this separation enables flexible and responsive design, it also makes editing and debugging CSS a challenging and time-consuming task, especially for users who are not familiar with how the DOM and CSS relate to each other. Even simple visual requests such as ``move this button to the center'' or ``change the background color of this section'' often requires understanding how multiple CSS rules interact for a given element, and how a change affects the rendered output, which makes CSS editing a trial and error process.

Existing approaches in web development debugging focus on three main directions. First, natural-language web styling and editing such as \textit{Stylette}~\cite{stylette} and \textit{Instruct4Edit}~\cite{instruct4edit} enable users to express visual changes in natural language, but they either rely on iterative suggestions or operate at the full page HTML modifications level rather than localized CSS edits. Second, visual debugging and failure repair techniques~\cite{WebSee, usingvisualsymptomsfordebugpresentationfailuresinwebapps, VLMcanvas} focus on detecting and localizing presentation failures by analyzing visual differences, often relying on reference renderings (oracles) and targeting fault diagnosis rather than learning from the correlation between CSS mutations (code changes) and its visual effect. Third, code generation from desired effects~\cite{divideandconquer} typically generate new HTML/CSS code structures from scratch instead of modifying on existing webpages.

In contrast, this work uses a mutation-based approach to model how natural language user intents relate to localized CSS changes applied to specific DOM elements, by learning from these mutations. The approach captures the mapping between code changes and user's natural language intent, making it applicable beyond failure scenarios to any arbitrary desired visual modifications.

To demonstrate the feasibility and practical impact of this approach, we conduct a case study on predicting website configuration changes through CSS mutations. Following a similar methodology as introduced in the approach \Cref{sec:approach}, we generate a large-scale dataset by mutating CSS rules on real website templates and capturing the resulting visual outcomes. Each mutation is paired with (1) a natural-language user intent, (2) the targeted DOM element and its original CSS code, and (3) before/after screenshots illustrating the visual effect of the change.

\begin{figure}[h]
    \centering
    \includegraphics[width=1\textwidth]{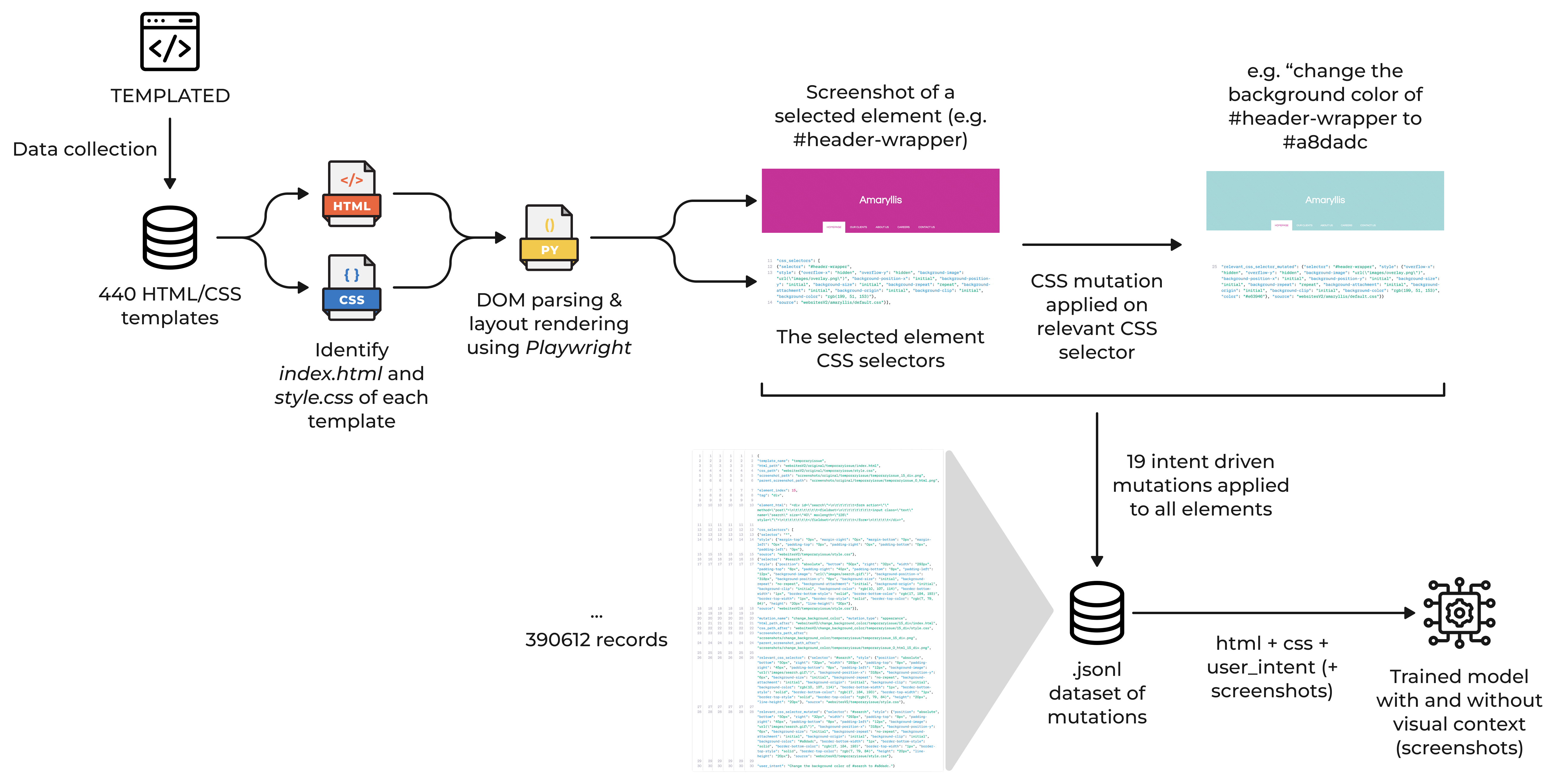}
    \caption{Overview of the intent driven CSS mutation pipeline}
    \label{fig:pipeline}
\end{figure}

\begin{figure}[h]
    \centering
    \includegraphics[width=0.8\textwidth]{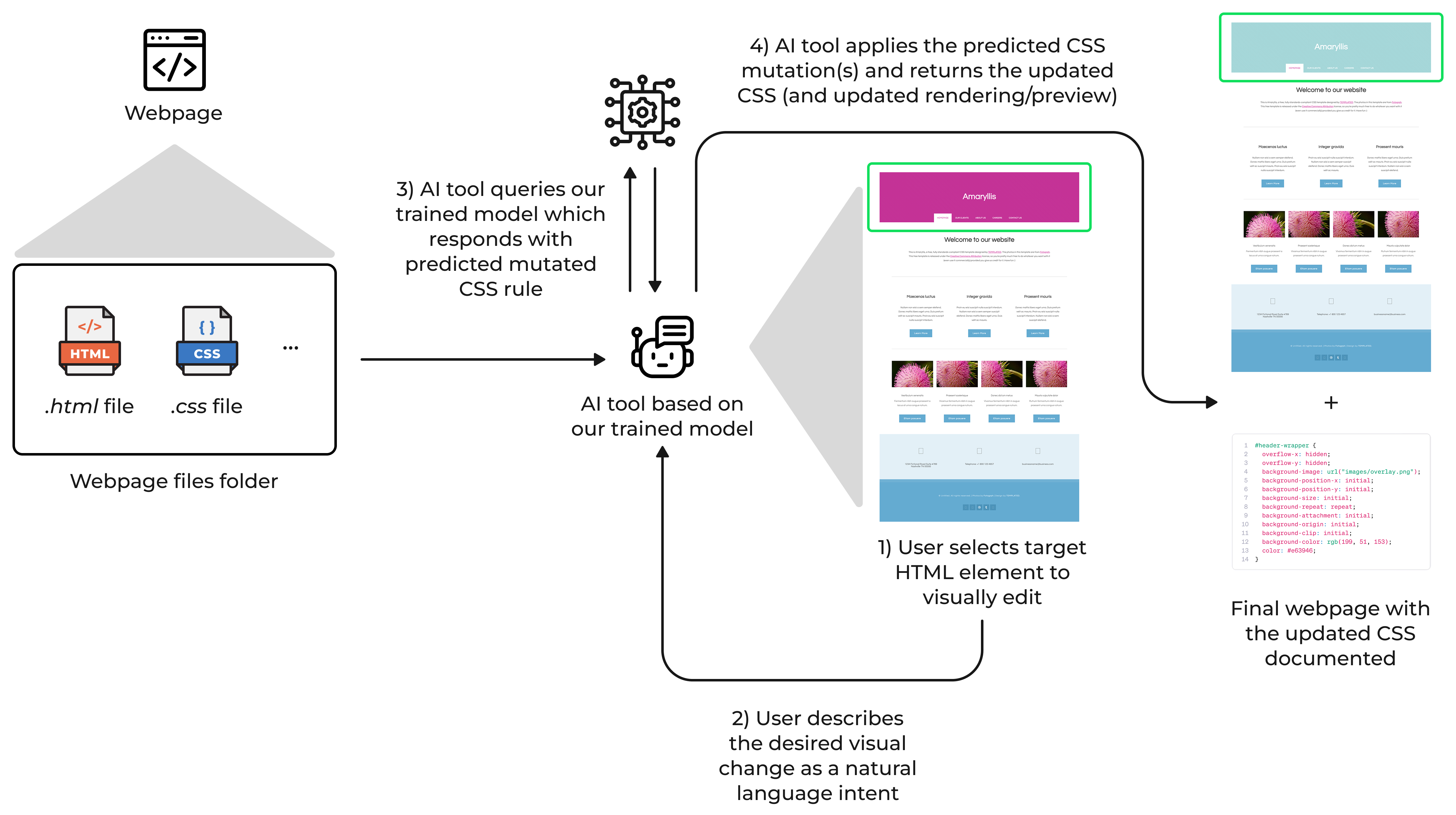}
    \caption{Example usage of the trained model as part of an AI tool for CSS editing}
    \label{fig:usage}
\end{figure}

To illustrate the overall workflow, \Cref{fig:pipeline} presents an overview of the proposed intent-driven CSS mutation pipeline, from dataset construction to model prediction. In addition, \Cref{fig:usage} shows an example usage of the trained model integrated into an AI-assisted CSS editing tool, demonstrating how user intents can be translated into concrete CSS changes in practice.

\subsection*{Dataset}

To construct a large-scale, diverse, and structured dataset of real-world website layouts whose HTML and CSS can be programatically analyzed and its CSS mutated, we follow these four steps:

\begin{enumerate}
    \item \emph{Collecting website templates.} We start by collecting 440 modern HTML/CSS website layouts from TEMPLATED\footnote{\url{https://templated.live}}, a public repository of free, open-source HTML5/CSS3 templates. For each template, we extract DOM structure and identify visible elements to serve as potential mutation targets.  
    \item \emph{Extracting element-level CSS rules.} For each DOM element, we collect the CSS rules that explicitly apply to it using the CSS Object Model (CSSOM) and selector matching. This allows us to preserve the relationship between the HTML element and its corresponding CSS selectors and blocks.
    \item \emph{Implementing \& applying CSS mutations.} We then apply different CSS mutations that simulate real-world user intents, including layout, appearance, and structure changes. Each mutation modifies a specific CSS rule according to a selector targeting logic. This step represents the key component of the approach as it allows gathering enough training samples covering a wide diversity of changes.
    \item \emph{Rendering \& screenshots collection.} For each mutated version of the webpage, we generate the before and after screenshots of the HTML element and its parent to capture the visual effect of the applied mutation.
\end{enumerate}
Total available dataset: 390612 samples.

\subsection*{Model Training and Application}

To evaluate the proposed approach we train and compare four models: \texttt{\gpt}, \texttt{gpt-oss-20b}, \texttt{CodeLlama-7B\-Instruct}, and \texttt{Qwen3-4B-Instruct}. To keep experiments comparable, all models are trained on the same samples, but the prompt format is adapted to the interface each model expects. The input consists of a natural-language \texttt{user\_intent} together with the HTML target element \texttt{element\_html} and its \texttt{relevant\_css\_selector}, and the output is the CSS edit corresponding to the mutation \texttt{relevant\_css\_selector\_mutated} in JSON format. Currently models are trained on subsets for cost and time reasons.

The evaluation is structured around three experiments: (1) The \emph{feasibility} of predicting configuration changes, (2) project specific training, and (3) training multimodal models to learn predicting website configuration changes.

\subsection{Experiment 1: Feasibility} 
In this experiment, we study whether models can learn to generate the correct CSS mutations from a natural language intent in a controlled setting. 
\subsubsection{Experimental Setup} To this end, we consider two representative subsets sampled from the full dataset, ensuring coverage of all mutation types: \textit{2k mutation records} to evaluate whether the approach is feasible, and \textit{20k mutation records} to evaluate whether performance improves when more training data is available. The data subsets are split into training, validation, and test sets using an \textit{80/10/10} ratio while preserving the mutation distribution across the splits.  

Performance is assessed using two metrics that differ in strictness: \texttt{Strict@} metric requiring an exact match of the predicted CSS block, and \texttt{Relaxed@} metric capturing whether the mutation-critical property/value change required by the intent is correctly applied.

Beyond this controlled setting, we also investigate how models handle more complex mutation scenarios. In particular, we focus on chained (swap) mutations, which require generating two coordinated CSS edits. Since these mutations are substantially harder and fail in the mixed setting, and due to their geometric nature, we isolate them by training a swap-only dataset and extend evaluation with additional metrics beyond exact matching. We measure property-level correctness for the geometric CSS properties: \texttt{left}, \texttt{top}, \texttt{width}, and \texttt{height} using both strict and relaxed criteria. The relaxed variant in this case considers predictions as correct if values fall within a tolerance range of $\pm 20px$. Furthermore, we compute IoU (Intersection over Union)\footnote{\url{https://viso.ai/computer-vision/intersection-over-union-iou/}} based on the predicted and expected bounding boxes to assess whether the resulting rendering matches the intended visual effect.

\subsubsection{Results} To establish a baseline, we first evaluate \texttt{\gpt} without fine-tuning on the 2k test set. The baseline achieves low performance, with $23.50\%$ under \texttt{Strict@} and $32.00\%$ under \texttt{Relaxed@}, indicating that a general purpose model cannot reliably predict the intended CSS mutations from user intent and DOM/CSS context alone. \Cref{tab:exp1.1-results} reports the strict and relaxed accuracy results across all models for both 2k and 20k training samples.

\begin{table}[H]
\centering
\caption{Strict@ and Relaxed@ results across all models for the 2k and 20k samples.}
\label{tab:exp1.1-results}
\begin{tabular}{lrrrr}
\rowcolor{black!10}
Dataset & \multicolumn{2}{c}{\textit{2k samples}} & \multicolumn{2}{c}{\textit{20k samples}} \\
\cline{2-5}
\rowcolor{black!10}
Metric & Strict@ & Relaxed@ & Strict@ & Relaxed@ \\
\noalign{\vskip 2pt}
\gpt{}  baseline          & 23.50\% & 32.00\% & - & - \\
\gpt                   & 51.50\% & 95.50\% & - & - \\
gpt-oss-20b                    & 46.50\% & 85.00\% & 47.00\% & 91.50\% \\
CodeLlama-7B-Instruct          & 43.50\% & 90.00\% & - & - \\
Qwen3-4B-Instruct              & 47.50\% & 90.50\% & 49.85\% & 92.90\% \\  
\end{tabular}
\end{table}

After fine-tuning on the 2k dataset, all models show substantial improvements, particularly under the \texttt{Relaxed@} metric. \texttt{\gpt} achieves the highest relaxed accuracy of $95.50\%$, while open-weight models also perform competitively, with \texttt{Qwen3-4B-Instruct} reaching $90.50\%$, \texttt{CodeLlama-7B-Instruct} $90.00\%$, and \texttt{gpt-oss-20b} $85.00\%$. This demonstrates that models can effectively learn to map natural-language intents to the corresponding CSS mutations, and that open-source models are particularly valuable in practice, as they can be fine-tuned and deployed locally, adapted to specific website contexts, and seamlessly integrated into development workflows without relying on commercial APIs.

Across all models, \texttt{Strict@} accuracy remains significantly lower, mainly due to exact matching constraints rather than true semantic errors. In many cases, models correctly apply the mutation-critical property/value changes but fail to reproduce the exact selector or full CSS block (e.g., different isolation suffixes or omitted unchanged properties). This explains the consistent gap between strict and relaxed performance and suggests that relaxed evaluation better reflects functional correctness in this task.

When increasing the training data from 2k to 20k samples, performance improves mainly under the \texttt{Relaxed@} metric, with \texttt{gpt-oss-20b} improving from $85.00\%$ to $91.50\%$ and \texttt{Qwen3-4B-Instruct} from $90.50\%$ to $92.90\%$. While, strict accuracy shows only marginal gains. This indicates that additional samples help models more consistently predict the mutation-critical properties changes.

Beyond atomic mutations, the swap only setting highlights the increased difficulty of chained mutations. When trained specifically on the swap only dataset, the model achieves moderate performance under relaxed and IoU-based evaluation, but struggles under strict matching. While predictions for \texttt{left}, \texttt{top}, and \texttt{width} are often close to the expected values, inconsistencies in jointly predicting both elements limit overall swap success.

\begin{table}[H]
\centering
\caption{Swap only samples evaluation using strict and relaxed accuracy and IoU rate.}
\label{tab:exp1.2-results}
\begin{tabular}{lrrr}
\rowcolor{black!10}
Property/Element & Strict@ & Relaxed@ & IoU@\\
\noalign{\vskip 2pt}
\texttt{left}   & 56.48\% & 72.22\% & - \\
\texttt{top}    & 50.00\% & 66.67\% & - \\
\texttt{width}  & 48.15\% & 78.70\% & - \\
\texttt{height} & 5.56\%  & 46.30\% & - \\
\texttt{element\_a} & - & - & 0.463 \\
\texttt{element\_b} & - & - & 0.611 \\
\end{tabular}
\end{table}

\highlight{Overall, these results confirm the feasibility of the proposed approach. Models can learn to generate correct CSS mutations from natural-language intents and DOM/CSS context, achieving up to $95.50\%$ relaxed accuracy. The large gap between \texttt{Strict@} and \texttt{Relaxed@} shows that many predictions are functionally correct even when they do not exactly syntactically match the expected CSS block. Scaling the training data further improves relaxed performance, while the swap only results highlight that chained mutations remain more challenging and require more precise coordinated geometry predictions across multiple CSS edits.}

\subsection{Experiment 2: Project-Specific Training} 

In this experiment, we study how models perform when training and testing are done \emph{on a single website template.} 
Unlike the cross-template setting, this setup keeps the HTML and CSS structure consistent across all samples. 

\subsubsection{Experimental Setup} To this end, we construct a template-specific dataset using the \texttt{premiumseries} template, which provides slightly more than \textit{3k samples} covering all mutation types. The dataset is split into training, validation, and test sets using the same \textit{80/10/10} ratio mentioned in \textit{Experiment 1}. The task formulation remains unchanged and the evaluation is performed using the same \texttt{Strict@} and \texttt{Relaxed@} metrics as in Experiment 1.

\subsubsection{Results} \Cref{tab:exp2-results} reports the \textit{Strict@} and \textit{Relaxed@} results for the project-specific setting using the \texttt{premiumseries} template (approximately 3k samples). Compared to the cross-template feasibility setting, all models show a clear improvement, particularly under the strict metric where up to 72.87\% are achieved compared to only 51.5\% by our general model from \Cref{tab:exp1.1-results}. Additionally, under the relaxed metric, 100\% accuracy were reached with the fine-tuned \gpt{} model, closely followed by the fine-tuned Qwen model with 95.9\% relaxed accuracy.

\begin{table}[H]
\centering
\caption{Project-specific strict and relaxed accuracy results on the \texttt{premiumseries} template (3126 samples)}
\label{tab:exp2-results}
\begin{tabular}{l c c}
\rowcolor{black!10} Dataset & \multicolumn{2}{c}{\textit{3k samples}} \\
\cline{2-3}
\rowcolor{black!10} Metric & Strict@ & Relaxed@ \\
\noalign{\vskip 2pt}
\gpt          & 72.87\% & 100.00\% \\
gpt-oss-20b           & 54.26\% & 91.80\% \\
CodeLlama-7B-Instruct & 54.89\% & 90.22\% \\
Qwen3-4B-Instruct     & 70.98\% & 95.90\% \\
\end{tabular}
\end{table}

The results show that all models benefit from the project-specific setting, with the improvement being particularly clear under the relaxed metric. \texttt{\gpt} reaches $100\%$ relaxed accuracy, indicating that when the template structure and styling remain consistent, the model can reliably generate visually correct CSS mutations. \texttt{Qwen3-4B-Instruct} also performs strongly on both strict and relaxed accuracy, showing that an open-source model can adapt well to a single-template setting. \texttt{CodeLlama-7B-Instruct} also performs competitively in this setting, reaching $90.22\%$ relaxed accuracy and $54.89\%$ strict accuracy, \texttt{gpt-oss-20b} also reaches more than $90\%$ relaxed accuracy. Under the strict metric, \texttt{\gpt} achieves $72.87\%$, \texttt{Qwen3-4B-Instruct} achieves $70.98\%$, and \texttt{gpt-oss-20b} achieves $54.26\%$, which are all higher than their corresponding strict results in Experiment 1.

Under the \texttt{Strict@} metric, most failures are caused by a mismatch in the automatically generated isolation class suffix. In these cases, the model typically outputs the correct CSS rule and the correct mutation critical change, but the selector string differs only in the isolation class. In contrast, results under the \texttt{Relaxed@} metric remain consistently high, showing that models reliably capture the intended visual changes. The remaining errors are mainly due to formatting issues or missing mutation-critical properties. Additionally, since chained (swap) mutations are largely absent in this setting, the results primarily reflect atomic mutations, which further contributes to the improved performance compared to the cross-template scenario.

\highlight{Overall, the higher strict and relaxed accuracies in this project-specific setting show that models successfully learn the CSS changes from user intents more reliably when training is focused on a single template with consistent HTML/CSS structure and conventions. Models achieve substantially higher \texttt{Strict@} and \texttt{Relaxed@} than in the cross-template setting, with \texttt{\gpt} reaching $100.00\%$ and \texttt{Qwen3-4B-Instruct} closely following with $95.90\%$. The remaining failures are mainly caused by selector class mismatches rather than unexpected visual changes.}

\subsection{Experiment 3: Multimodal Model With Screenshot Context} 
In this experiment, we study the impact of adding visual context on the models' ability to generate the intended CSS mutations. 

\subsubsection{Experimental Setup} While previous settings rely solely on code and textual input, here we extend each training instance with screenshots of the target element and its container layout or parent element captured both before and after the mutation. We train a vision-language model (\texttt{Qwen2-VL-2B-Instruct}) using the same dataset construction and evaluation protocol as in previous experiments. The screenshot inputs are combined with the same previous code/text inputs (user intent, element HTML, and CSS rule), and the model is trained to produce the mutated CSS block in structured JSON format. Results are reported using the same \texttt{Strict@} and \texttt{Relaxed@} metrics to ensure comparability, and experiments are conducted on both smaller (2k samples) and larger (20k samples) training subsets to assess how the contribution of visual context evolves with more data. 

\subsubsection{Results} \Cref{tab:exp3-results} reports the \textit{Strict@} and \textit{Relaxed@} results for the multimodal setting using \texttt{Qwen2-VL-2B-Instruct}. Compared to the code/text only model, adding screenshot context leads to consistent improvements, particularly under the \texttt{Relaxed@} metric.

\begin{table}[H]
\centering
\caption{RQ2 Multimodel training with screenshot context strict and relaxed accuracy results on both the 2k and 20k samples.}
\label{tab:exp3-results}
\begin{tabular}{l c c c c}
\rowcolor{black!10} Dataset & \multicolumn{2}{c}{\textit{2k samples}} & \multicolumn{2}{c}{\textit{20k samples}} \\
\cline{2-5}
\rowcolor{black!10} Metric & Strict@ & Relaxed@ & Strict@ & Relaxed@ \\
\noalign{\vskip 2pt}
Qwen2-VL-2B-Instruct     & 47.50\% & 93.50\% & 54.35\% & 97.10\% \\
Qwen3-4B-Instruct              & 47.50\% & 90.50\% & 49.85\% & 92.90\% \\
\end{tabular}
\end{table}

At 2k samples, the multimodal model achieves $93.50\%$ relaxed accuracy compared to $90.50\%$, while strict accuracy remains unchanged at $47.50\%$. At 20k samples, the improvement becomes more pronounced, with $97.10\%$ relaxed accuracy and $54.35\%$ strict accuracy, compared to $92.90\%$ and $49.85\%$ for the code/text only model.

As in previous experiments, strict performance is mainly limited by exact matching constraints, with most failures caused by minor differences in the generated selectors rather than incorrect mutation logic. Under the relaxed metric, the multimodal setting reduces the number of atomic mutation errors, indicating that visual context helps the model better capture the intended visual changes. However, remaining errors are still concentrated in more complex cases such as chained (swap) mutations, suggesting that screenshot context alone does not fully resolve these challenges.

\highlight{Overall, these results show that incorporating screenshot context has a clear positive impact on learning CSS mutations from user intent, especially as training scales. Notably, the model achieves up to $97.10\%$ relaxed accuracy with a relatively small open-source vision-language model, highlighting the effectiveness and practical applicability of the approach in real-world web development scenarios.}

\subsection{Summary of Case Study Findings}

Across all experiments, the results demonstrate that models can effectively learn to predict CSS mutations given a natural language user intent. While strict accuracy is consistently limited by exact matching constraints, relaxed accuracy shows that models reliably capture the intended visual changes. Performance further improves when training is conducted in more controlled settings, such as a project-specific scenario, and when additional visual context is provided through screenshots. While more complex mutation types, such as chained (swap) mutations, remain challenging due to the need to predict consistent geometry and maintain coordinated mutations across two elements. Overall, the findings highlight the feasibility and practical potential of intent-driven CSS mutations, especially when using open-weight and multimodal models.

%% file: sections/experiments.tex
\section{Evaluation}
\label{sec:experimental_evaluation}

The following sections will present the research questions, introduce the datasets used in this study, the applied evaluation metrics, and the experimental setup and results for each experiment.

\subsection*{Research Questions} 

To investigate the feasibility of \emph{predicting software changes from desired behavioral outcomes}, 
we focus on the following research questions (RQs). 
These questions guide our evaluation of the approach, and lead to the creation of several datasets and to the definition of the evaluation metrics.

While RQ1 investigates the feasibility of Neural Change Prediction, RQ2--RQ6 investigate the factors that influence the performance of neural change prediction.

\begin{description}
    \item[RQ1:] \textbf{To what extent can the effects of software changes be learned to predict code changes based on desired behavioral outcomes?}
    This question examines the predictive power of our models in linking software changes with behavioral outcomes. 
    In particular, we investigate the ability of the models to 
    (i)~localize where in the code a change should occur to achieve a specific behavioral modification, 
    (ii)~generate the concrete code change required, and 
    (iii)~predict the effect of a given code change on program behavior.

    \item[RQ2:] \textbf{How many generation attempts are required on average to successfully predict neural changes?} Machine learning models are typically non-deterministic (unless temperature set to force determinism), thus it is important to assess how many attempts are required on average to obtain the desired output from the model while optimizing resources. 

    \item[RQ3:] \textbf{How accurately can behavior changes leading to failures be detected, and the introduced errors predicted?} In this question we will not only investigate behavior changes that alter the programs output but also investigate code changes that introduce errors or timeouts. A detailed analysis on how well different types of errors can be predicted from code changes will be performed.

    \item[RQ4:] \textbf{Does the input complexity of the program of interest impacts the performance of Neural Change Prediction?} Program inputs can be standard types or objects like graphs. In this experiment, we investigate whether the complexity of the input influences the learning of how code and behavior changes relate.

    \item[RQ5:] \textbf{How well does Neural Change Prediction scale from Single Order Mutants to Higher Order Mutants?} In RQ1 the feasibility of learning the effects of single order software changes has been demonstrated. To scale this to more complex changes, and demonstrate the applicability of this work, we investigate how well Neural Change Prediction performs on higher order changes.

    \item[RQ6:] \textbf{How does more context about the programs' dynamic behavior influence the models capabilities in learning to predict code changes?} Representing a program's behavior in natural language, not only for a single input, but in general is a challenging task. Thus, we extended the learning context to include all input-output pairs from the test set. Comparing the performance of models fine-tuned with different behavior context provides insights into the current performance and motivate future work directions. 
\end{description}

\subsection*{Datasets}

For our experiments, we used the open-source Python projects from the QuixBugs~\cite{lin2017quixbugs} benchmark which was selected for its diversity, size and complexity as well as its wide adoption to evaluate code generation tasks. QuixBugs contains a total of 40 projects, we will distinguish between 31 simple projects from QuixBugs and 9 projects from QuixBugs with more complex inputs, as detailed in \Cref{tab:datasets_py}. The projects complexity mainly differs in the type of inputs that the program is consuming, for ``simple'' the program takes standard input types like, e.g., int, string,list, etc. while ``complex'' programs consume more complex objects. Thus, the projects were split by input complexity for the experiments. In total we considered 44 functions that were mutated with an original average size of 13 lines of code per function.

To collect a large amount of (code change -- behavior change) pairs, we used PyMut4SE\footnote{\url{https://github.com/LaPlei96/PyMut4SE}}.
The mutation tool takes a software project as input, and for each Python function, it generates its Abstract Syntax Tree (AST) and applies all compatible mutation operators that can be applied to each node of the AST to produce program mutants. Taking inspiration from traditional mutation testing operators~\cite{king1991fortran,5487526} performing logical, relational, arithmetic and unary transformations to the AST as well as mutation operators changing the control flow of Python programs by changing or deleting conditions as done by the MutPy~\cite{derezinska2014analysis} tool.
In this study, in addition to the previously mentioned operators, PyMut4SE implemented type cast mutations, changes in variable assignments as well as changing the mandatory, default and optional parameters in function calls. 
More specifically, the mutation operators were applied to produce single order mutants (SOM) as well as high order mutants (HOM) to produce mutants of varying complexity.
HOMs produce more complex behavioral changes and increase the richness of our dataset.

Our database contains the original projects and their execution data as well as the mutants with metadata. For every code chunk (original and mutated), we collected the following: code, function name, mutation degree, applied mutation operator, original code, parent code, modified location, and path to project. For each execution, the execution environment, input, output, success state, potential error messages and further logs were stored in the database. 
By running our mutation tool on all QuixBugs subjects, we had to limit the maximum mutation degree to avoid an uncontrolled explosion of the amount of mutants. A total of 3,004,715 
executions from mutants of varying degrees were collected.

Throughout the experiments, different datasets introduced in \Cref{tab:datasets_py} were used to answer the research questions. While creating the datasets, equivalent mutants were omitted as they do not contribute to learning the effect of software changes.

\begin{table}[t] 
    \centering
    \caption[Python Datasets]{Python Datasets. \textit{SOM} stands for single order mutants, \textit{HOM} for higher order mutants, \textit{``Simple''} means that the inputs were from standard data types while \textit{``Complex''} represent specific objects as inputs, \textit{``Success''} refers to mutants whos' execution was successful while \textit{``All''} includes the mutants that produced errors during execution---including the exact error, \textit{Pairs} refers to including all input-output pairs.}
    \label{tab:datasets_py}
    \begin{tabular}{lr}
    \rowcolor{black!10}
        Dataset &  Samples\\
        \noalign{\vskip 2pt}
        SOM\_Simple\_Success & 2065\\
        SOM\_Simple\_All & 5241\\
        SOM\_Simple\_Success\_Pairs & 430\\
        SOM\_Complex\_Success & 64\\
        SOM\_Complex\_All  & 395\\
        SOM\_Complex\_Success\_Pairs & 10\\
        HOM\_Simple\_Success & 74158\\
        HOM\_Simple\_Success\_Pairs & 29640\\
        HOM\_Complex\_Success & 1201\\
        HOM\_Complex\_Success\_Pairs & 705\\
    \end{tabular}
\end{table}

\subsection*{Model}
In our experiments, we fine-tune the \texttt{gpt-4.1-mini-2025-04-14} model from OpenAI.
At the time of submission, the more recent GPT models were not yet available for fine-tuning.
Nevertheless, our approach does not depend on a specific model architecture and could easily be repeated with the rise of new models.
The aim of this work is to introduce Neural Change Prediction as a new way of learning the effects of software changes and generating the required code changes.
The focus was not to perform an in depth analysis of the currently available commercial and open-source LLMs.
We chose the OpenAI model was as it provides a good trade-off between performance and resources. 
For the fine-tuning, the OpenAI API was used with the default configuration. In each experiment, the dataset was split into 80\% for training, 10\% for validation and 10\% for testing.

\subsection*{Evaluation Metrics}\label{eval_metrics}

We evaluated the models across the three target tasks:

\begin{description}
    \item[Predicting Locations.] We measured the accuracy of identifying the correct mutation location $L$. For SOM there is only a single location that starts at line $L$ while in HOM there are several locations that were changed. HOM change locations are represented as a list of $L$, starting with the line that was modified by the first mutation until the line changed by the last mutation. Thus, for HOM we will distinguish between:
    \begin{enumerate}
        \item \textbf{One Location.} We check if at least one of the locations that needs to be modified has been detected. 
        \item \textbf{First Location.} We evaluate if the first location to modify has been correctly identified.
        \item \textbf{Set Locations.} This metric is true, only if all locations have been correctly predicted, independent of their modification order.
        \item \textbf{All Locations Strict.} This metric is true, only if all locations match the expected locations in the correct order of changes. 
    \end{enumerate}
    \item[Predicting Effects.] We check whether a given code change produces the desired output $o'$.
    \item[Generating Code Changes.] We evaluated two complementary binary metrics:
    \begin{enumerate}
        \item \textbf{Identical Code:} set to $1$ if the generated code is syntactically identical to the expected code and $0$ otherwise.
        \item \textbf{Semantic Code Clone:} set to $1$ if, when executing the generated and expected code on all test inputs, executions on each input produced identical outputs for both code versions; otherwise $0$.
    \end{enumerate}
\end{description}

\subsection{RQ1: Learning the Effects of Software Changes}\label{sec:rq1}

\emph{To what extent can the effects of software changes be learned to predict code changes based on desired behavioral outcomes?}\
To address RQ1, we investigate the feasibility of our models to: 
(1)~localize where in the code, a change should occur to achieve a specific behavioral modification;
(2)~generate the required code change; and 
(3)~predict the effect of a given code change on program behavior. 
We detail the experimental setup, the used dataset, and the training of the model. 
We then present and discuss the results.

\subsubsection{Experimental Setup}
To investigate the feasibility of Neural Change Prediction, we used the SOM\_Simple\_Success dataset from \Cref{tab:datasets_py}. The dataset is composed of single order mutants that were successfully executed. For all three tasks, the dataset was formatted accordingly to the data relevant for the task, more specifically, the input and output features where chosen as described in \Cref{fig:workflow}. We fine-tuned the \gpt{} model for each task, leveraging the models syntactic knowledge of Python code, while providing the dynamic knowledge of the projects by adding the semantic signals observed at runtime to learn from. 
The dataset was split into 80\% for training, 10\% for validation, and 10\% for testing.
Fine-tuning required preparing the training and validation data as \texttt{.jsonl} files, where each entry consists of a natural language prompt and the expected output. 
An example of such a prompt for the task of code generation is illustrated in~\Cref{lst:prompt}.\\

\begin{lstlisting}[language=json,caption=Prompt template for the code generation task,label=lst:prompt,float=t,captionpos=b,showspaces=false,showstringspaces=false]
{
  "messages": [
    {
      "role": "system",
      "content": "Be a helpful software engineer assistant."
    },
    {
      "role": "user",
      "content": "I have the following Python code: {ORIGINAL FUNCTION F} and for this {INPUT i} it gives me the following {OUTPUT O} But I want this {OUTPUT O'} Please write me the corresponding Python code"
    },
    {
      "role": "assistant",
      "content": "Here is the corresponding Python code: {CHANGED FUNCTION F'}"
    }
  ]
}
\end{lstlisting}

To assess the performance of the fine-tuned models on our test set, a baseline was established first by prompting the \gpt{} base models with our test samples.

\subsubsection{Results}\label{resultsrq1}

We present the results of evaluating our fine-tuned \gpt{} model 
on the three core tasks: localization, code change generation, and behavior change prediction. 
\Cref{tab:rq1results} summarizes our results in comparison to the baseline results.

\begin{table}[h]
  \caption{Accuracy of fine-tuned \gpt{} on SOM\_Simple\_Success for all three tasks vs baseline}
  \label{tab:rq1results}
  \begin{tabular}{lrr}
    \rowcolor{black!10}
    Task & Baseline \gpt{} & Fine-tuned \gpt{} \\
    \noalign{\vskip 2pt}
        Localization & 9.6\%  & 82.6\%\\[5pt]
        Code Change Generation - Identical Code &  0.0\% & 51.1\%\\
        Code Change Generation - Semantic Code Clone &  12.8\% & 68.5\%\\[5pt]
        Behavior Change Prediction &  33.3\%  & 87.1\%\\
\end{tabular}
\end{table}

\noindent
\textbf{Localization.} LLMs have seen a huge amount of code and natural language descriptions which allows them to map natural language description to what the code is supposed to do~\cite{shi2025natural}, thus they are able to perform some minimalistic static analysis. However, since they lack the dynamic information about software projects, they struggle to relate which part of the code affects specific behaviors of the software as can be seen with the baseline only reaching 9.6\% for the localization task. Eventhough the dataset in this experiment only contained single order mutations, the model could not detect which part had to be changed to achieve the desired behavior. After fine-tuning, 82.6\% accuracy were reached for fine-grained localization. Only a single test sample was predicted correctly by the baseline but not by our model. This significant improvement highlights the models capabilities of learning the effects of software changes and where they were made. Those promising results highlight as well the need to learn from executions to gather dynamic information about a project as well as for Neural Change Prediction to target more complex higher order changes as we will investigate in \Cref{sec:rq2}.

\highlight{Neural Change Prediction allows to accurately predict where to apply a change to achieve the desired effect, the need of learning from software changes has been highlighted by an accuracy increase from 9.6\% up to 82.6\%.}

\textbf{Code Change Generation.} Generating code that exactly matches the desired behavior is a challenging task. Often, generated code behaves similarly to the developer's intended code but fails to achieve the desired behavior in all execution environments on a diversity of inputs. In this study, we aim at generating code that is semantically equivalent to the expected code, where semantic equivalence is assessed by running the generated code in a dedicated execution environment on all the inputs from the test suite (i.e., if the output is the same for all inputs of the test suite, we consider semantic equivalence). By assessing the baseline capabilities of the \gpt{} model, only 12.8\% of the generated code were semantically equivalent to the desired code with no exact match.
Our \emph{fine-tuned model} achieved an accuracy of 68.5\% in generating code that is semantically equivalent to the expected code, demonstrating the benefits of learning from pairs of code and behavior changes.
    The relatively small difference between semantic code clones and identical code can be explained by us performing this experiment on SOM.
    Thus, there is only a relatively small number of semantic code clones that achieve the same behavior.

\highlight{Code generation is the most complex task for Neural Change Prediction, nevertheless our model reached an accuracy of 68.5\% in generating semantic code clones, a significant improvement from the baseline.}

\textbf{Behavior Change Prediction.} By default, on our test set, \gpt{} was able to correctly predict the effect of software changes in 33.3\% of the cases. From documented code samples, LLMs have gained some knowledge of how code snippets behave on simple inputs, which allowed the model to achieve the best performance across all three tasks for the effect prediction. Nevertheless, the dynamic knowledge of a project is still essential to learn from, as the results from the fine-tuned model demonstrate. With fine-tuning, the model is capable of learning meaningful relationships between code changes and program behavior, leading to an accuracy of 87.1\%.

\highlight{Neural Change Prediction is not only feasible but allows to accurately predict the effect of a specific change to the software with 87.1\% accuracy. Overall, Neural Change Prediction can significantly improve models that assist developers for software engineering tasks.}

\subsection{RQ2: Number of Generation Attempts}\label{sec:rq2}

\emph{How many generation attempts are required on average to successfully predict neural changes?}
Large Language Models are known to be non-deterministic, varying their response when repeatably queried with the same prompt (unless temperature set to force determinism). Developers ideally want to optimize the generation attempts to reduce the cost and time required to get the desired code or behavior change. Additionally, models generating different suggestions could lead to a developer investigating several options. Studying how many attempts are typically required and performing majority voting across the generated suggestions can significantly reduce the developers investigations.
It is thus essential to extend the analysis of the \gpt{} model fine-tuned in \Cref{sec:rq1} on the SOM\_Simple\_Success dataset for all three tasks.

\subsubsection{Experimental Setup}\label{exp_gen_attempts}
We consider the \gpt{} fine-tuned on SOM\_Simple\_Success for all three tasks from RQ1 and further evaluated the model while performing only one versus several (maximum five) generation attempts. The accuracies obtained with up to five generations and the average attempts required to achieve a successful response are reported and discussed.

\subsubsection{Results}

\begin{table}[b]
  \caption{Accuracy of fine-tuned \gpt{} on SOM\_Simple\_Success for all three tasks. We report the accuracy with a single generation attempt, maximum five generation attempts and the average attempts needed for successful prediction.}
  \label{tab:rq2genattempts}
  \begin{tabular}{lrrr}
    \rowcolor{black!10}
    Task & Single Attempt & Multiple Attempts & average \# attempts\\
    \noalign{\vskip 2pt}
        Localization & 66.0\%  & 82.6\% & 1.40\\[5pt]
        Code Change Generation - Identical Code &  26.0\% & 51.1\% & 1.91\\
        Code Change Generation - Semantic Code Clone &  46.1\% & 68.5\% & 1.56\\[5pt]
        Behavior Change Prediction &  84.3\%  & 87.1\% & 1.04\\
\end{tabular}
\end{table}

As can be seen in \Cref{tab:rq2genattempts}, the accuracy increases across all three tasks with more generation attempts. We performed up to five generation attempts unless an identical code was generated earlier. However, the average number of attempts required to achieve a correct prediction/generation changes significantly between tasks. Generating the correct code that is identical to the expected code requires almost twice as many attempts as predicting the behavior change given a code change. The average number of required generation attempts can guide the amount of generations performed in practice by the developer. For behavior change prediction, no correct prediction required more than two attempts, for semantic code clones and localization five attempts were sufficient while for the generation of identical codes, the maximum number of attempts could be slightly increased since 8\% of the successfully generated identical codes were generated during the fifth attempt. However, in practice, the semantic code clones property should be the aim when requiring code changes to a achieve a desired behavior. Thus, a developer can use Neural Change Prediction to make one or more suggestions on how to change the code. With the knowledge of the average required attempts for each task, a developer can optimize the number of attempts to only get a small set of changed programs. Since changed programs can always be executed, it can always be verified that the desired change actually occurred.

\highlight{On average between one and two generation attempts are required for correct predictions with code generation requiring the most generation attempts, especially to obtain code changes identical to the expected code. For behavior change prediction, the fine-tuned model is very confident and a single generation attempt is overall sufficient.}

\subsection{RQ3: Behavior Change Prediction in Successful and Failing Cases}\label{sec:rq3}

\emph{How accurately can behavior changes leading to failures be detected, and the introduced errors predicted?}
Predicting how the behavior of software changes can be useful in different scenarios.
The software might be working as intended and modified to add new functionalities, change a library, upgrade to a newer version or incorporate new components.
In that case, a code change might not introduce any side effects and the behavior remains the same or the behavior is altered.
While programs can run successfully but behave differently than expected by the developer, in that case a change in behavior can be expressed, e.g., by the desired change in a programs output. 
However, frequently simple code changes can also lead to different kinds of errors or crashes, not just altering the software's output.
Thus, predicting the effect of software changes before integrating modifications into a system is essential.

Nevertheless, behavior changes can also be very subtle and only visible in very specific cases or even only for a single input. 
In such cases, it is important to learn from software changes that might induce or fix those errors.
Pairs of code changes where one produces an output and the other throws an error are important in the context of code generation (in this case automated repair) as well as in the context of effect prediction, to prevent a code change with an undesired side effect.

\subsubsection{Experimental Setup}

In addition to the first experiment discussed in \Cref{sec:rq1} where we predicted behavior changes for \emph{successful} executions only, we now assess the feasibility of detecting and predicting \emph{error types.}
To that end, we fine-tuned the \gpt{} model on the SOM\_Simple\_All dataset which includes samples of successful and failing executions.
In the event of a successful execution, the output of the function was part of the prompt. In the event of a failure, if an error was thrown, the error type was included in the prompt.
In the case of a timeout, the prompt contained the information about the timeout.
Hence, the fine-tuned model can predict whether the code change will produce the desired behavior change or an error.

\subsubsection{Results}
As shown in \Cref{tab:rq2success}, the fine-tuned model was able to accurately predict 95\% of all behavior changes. When evaluating only the code changes that introduced an error or timeout, even 99\% were correctly detected by our fine-tuned model. From the results of the model previously (\Cref{sec:rq1}) fine-tuned on the SOM\_Simple\_Success dataset, it seems that predicting the effect of software changes that introduce severe side effects is easier than predicting more specific software changes. Nevertheless, our first model is still able to correctly predict the minor effects of software changes with 87.1\% accuracy. Both models fine-tuned to predict the effect of software changes are very confident in their predictions as they only required 1.04 attempts on average on the SOM\_Simple\_Success and even only 1.03 on the SOM\_Simple\_All dataset.

\begin{table}[t]
  \caption{Accuracy of fine-tuned \gpt{} on SOM\_Simple\_Success and on SOM\_Simple\_All}
  \label{tab:rq2success}
  \begin{tabular}{lrr}
    \rowcolor{black!10}
    Task & SOM Success only & SOM all \\
    \noalign{\vskip 2pt}
        Behavior Change Prediction & 87.1\% & 95.0\% \\
\end{tabular}
\end{table}

In addition, we assessed which error types are the easiest to learn and predict. \Cref{tab:rq2errors} highlights that the introduction of IndexError and OverflowError in software changes were the hardest to detect while overall, the models predictions reached 100\% accuracy for almost all error types. The poor reported performance for OverflowError might simply be due to their scarcity in the dataset, in total only four OverflowErrors were present in the full dataset, with only one as test sample, thus the results for the OverFlowError might not be representative for the feasibility of learning to predict such errors. Additionally, the model predicted in a few cases that an error would occur even-though the code change was producing the desired behavior change. Nevertheless, this only occurred in a few isolated cases as can be deduced from the precision reported in \Cref{tab:rq2errors}.

\begin{table}[h!]
  \caption[Performance of fine-tuned \gpt{} on predicting Error Types on SOM\_Simple\_All]{Performance of fine-tuned \gpt{} on predicting Error Types on SOM\_Simple\_All. \textit{Recall} describes how many errors of that type were correctly predicted while \textit{Precision} describes how many errors of each kind were predicted that were actually correct.}
  \label{tab:rq2errors}
  \begin{tabular}{lrr}
    \rowcolor{black!10}
    Error Type (samples in test set) & Recall & Precision \\
    \noalign{\vskip 2pt}
        AttributeError (16)& 100\%  & 100\% \\
        IndentationError (16)& 100\%  & 100\% \\
        IndexError (23) & 91\%  & 95.7\% \\
        KeyError (19)& 100\%  & 100\% \\
        NameError (28) & 100\%  & 100\% \\
        OverflowError (1) & 0\%  & 0\% \\
        RecursionError(26)	 & 100\%  & 92.9\% \\
        TypeError (155) & 100\%  & 100\% \\
        UnboundLocalError (6)& 100\%  & 100\% \\
        ValueError (13)& 100\%  & 100\% \\
        ZeroDivisionError (10) & 100\%  & 90.9\% \\
        Time out (19) & 100\%  & 90.5\% \\

\end{tabular}
\end{table}

Overall, predicting the effects of software changes can be effectively learned from pairs of code and behavior changes to assist developers while debugging, maintaining, and extending their software.

\highlight{Detecting software changes that introduce errors or timeouts can successfully be detected with very high recall and precision. While predicting error types given a code change is easier than predicting a subtle behavior change, even these can be correctly predicted with 87.1\% accuracy.}

\subsection{RQ4: Input Complexity}\label{sec:rq4}

\emph{Does the input complexity of the program of interest impact the performance of Neural Change Prediction?}

The complexity of projects and learning from their executions does not only depend on code complexity but also on input complexity. While it might seem straightforward how the value of an input like an Integer might vary throughout the program's execution, it seems harder to assess the impact of software changes when the input is an object.
Some challenges in learning the effect of software changes executed on complex inputs come from the input representation.

\subsubsection{Experimental Setup}
The QuixBugs benchmark used in this study contains projects of different complexity. They particularly differ in their input type; two thirds of the projects take simple inputs such as int, string, list, etc. while one third of projects take complex inputs, e.g., objects like graphs. To successfully apply Neural Change Prediction on more complex inputs, a textual representation of the input was required. To address this challenge, in this study, we provided not only the name of the actual input (e.g. ``length\_by\_edge, node0, node1'') but also the textual representation of how to construct the object as illustrated in Listing~\ref{lst:input_rep}. In a series of experiments, we will compare the performance of the models on samples with simple and complex inputs.
Thus, we fine-tuned \gpt{} for all three tasks on the SOM\_Complex\_Success dataset. To validate our findings of RQ3~\Cref{sec:rq3}, we also fine-tuned \gpt{} on the SOM\_Complex\_All dataset.

\begin{minipage}{\linewidth}
\begin{lstlisting}[language=Python, caption=Example of input representation for objects, label=lst:input_rep,     basicstyle=\ttfamily\small,columns=fullflexible,showspaces=false,showstringspaces=false]
""" Given the following function definition: """
def shortest_path_length(length_by_edge, startnode, goalnode):
    ...

""" A possible function call would be: """
shortest_path_length(length_by_edge, node0, node1)

""" Input representation: """
{'representation': '\n'
 'node1 = Node("1")\n'
 'node5 = Node("5")\n'
 'node4 = Node("4", None, [node5])\n'
 'node3 = Node("3", None, [node4])\n'
 'node2 = Node("2", None, [node1, node3, node4])\n'
 'node0 = Node("0", None, [node2, node5])\n'
 'length_by_edge = {(node0, node2): 3,\n'
 '(node0, node5): 10,\n'
 '(node2, node1): 1,\n'
 '(node2, node3): 2,\n'
 '(node2, node4): 4,\n'
 '(node3, node4): 1,\n'
 '(node4, node5): 1,}\n'
 'return (length_by_edge, node0, node1)'}
\end{lstlisting}
\end{minipage}

\subsubsection{Results}
\Cref{tab:inputcomplexity} shows the performance of the fine-tuned model on projects with complex inputs, compared to the same base model trained on projects with simple inputs (RQ1~\Cref{sec:rq1} and RQ2~\Cref{sec:rq2}), to keep the comparison sound, up to five generation attempts were performed. 

\begin{table}[!h]
    \caption{Accuracy of fine-tuned \gpt{} on Projects with simple vs complex program inputs}
    \label{tab:inputcomplexity}
    \centering
    \begin{tabular}{lrr}
    \rowcolor{black!10}
        Task & SOM\_Simple\_Success & SOM\_Complex\_Success \\
        \noalign{\vskip 2pt}
      Localization   & 82.6\% & 76.9\% \\[5pt] 
      
      Code Change Generation - Identical Code & 51.1\% & 71.6\%\\
      Code Change Generation - Semantic Code Clone & 68.5\% & 71.6\%\\[5pt]
      
      Behavior Change Prediction & 87.1\% & 80.0\% \\
     
    \end{tabular}
\end{table}

Overall, the model fine-tuned on the SOM\_Complex\_All dataset reaches similar performance on all three tasks, with variations following the same trend as for the model trained on the SOM\_Simple\_Success dataset. Behavior change prediction remains the easiest task, where the model still achieves 80\% accuracy even with complex inputs and outputs. This highlights that Neural Change Prediction can be applied to learn from complex input and output samples as long as the textual representation of the input construction is provided. Localizing where to change code to achieve desired changes is slightly harder with more complex inputs. Interestingly, for code change generation, no difference was observed between semantic code clones and identical codes, meaning that all successful code change generations were identical to the expected code. Since for this experiment only the successful executions were kept, the resulting SOM\_Complex\_Success dataset is small and might represent a threat to validity. Many simple mutation operators lead to mutants that could not successfully be executed on the complex inputs. Nevertheless, those samples were kept to investigate the feasibility of learning the behavioral changes from those. As can be seen in \Cref{tab:rq2successcomplex}, it is feasible to accurately predict the effect of software changes across the SOM\_Complex\_All dataset with 92.3\% accuracy.

\begin{table}[!h]
  \caption{Accuracy of fine-tuned \gpt{} on SOM\_Simple\_Success and on SOM\_Complex\_All}
  \label{tab:rq2successcomplex}
  \begin{tabular}{lrr}
\rowcolor{black!10}
    Task & SOM Success only & SOM all\\
    \noalign{\vskip 2pt}

        Behavior Change Prediction & 80.0\% & 92.3\% \\
  
\end{tabular}
\end{table}

While learning from successful and failing executions on projects with complex inputs, we observed that predicting the effect of software changes that introduce failures is easier than predicting subtle behavior changes, similar to the experiments with the simple inputs reported in \Cref{tab:rq2success}.

\highlight{Neural Change Prediction can not only be applied on programs with simple inputs but even on complex inputs. This highlights the models capabilities in learning the relationship between program-input pairs and the resulting behavior as well as learning the effect of software changes for all kinds of behavior changes.}

\subsection{RQ5: Single Order Mutants vs Higher Order Mutants}\label{sec:rq5}

\emph{How well does Neural Change Prediction scale from Single Order Mutants to Higher Order Mutants?}
Real world software changes typically target several code locations where more than one change must be applied. While changing a single line often already influences the behavior of a software, more complex behavioral changes might need a sequence of changes referred to as higher order mutants, thus it is important to assess how well models can learn single changes versus complex changes required to achieve a desired behavior.

\subsubsection{Experimental Setup}
We conducted an experiment to assess the feasibility of Neural Change Prediction to learn from higher order mutants. For this purpose, the datasets HOM\_Simple\_Success and HOM\_Complex\_Success were created. In both, we omitted equivalent mutants, which do not help to learn the observable effect of software changes. Additionally, only successfully executed mutants were kept. Furthermore, the number of mutants whose execution resulted in null or infinity was significantly reduced in the dataset. More precisely, only one sample per project name and input id pair was kept. Since the resulting dataset was still very large for HOM\_Simple\_Success, we only kept the following samples: 
\begin{enumerate}
    \item All degree 2 mutants that were successfully executed (38,026 samples for HOM\_Simple\_Success). 
    \item For every project in QuixBugs a maximum of 2,500 mutants of degree 3 (36,132 for HOM\_Simple\_Success).
\end{enumerate}
Setting the threshold to 2,500 mutants of degree 3 per project allowed for an almost equal sample size for degree 2 and degree 3. A balanced dataset being crucial for successful learning and evaluating the models capabilities to predict where and what changes to make, as well as predicting the effect of software changes.

For code generation and effect prediction, the same evaluation metrics as in previous experiments will be applied. To evaluate the localization, we will distinguish between three localization metrics of varying granularity as introduced in \Cref{eval_metrics}.

To allow for a fair comparison of how well a model can learn the effect of more complex software changes, \gpt{} was fine-tuned once on HOM\_Simple\_Success and once on HOM\_Complex\_Success.

\subsubsection{Results}

Based on the findings from experiment 1 (\Cref{exp_gen_attempts}) on the generation attempts, only up to two generation attempts were performed for the task of behavior change prediction and up to three for the tasks of code generation and  for the HOM evaluation. All the results are reported in \Cref{tab:rq2hom}.

\begin{table}[b]
  \caption{Accuracy of fine-tuned \gpt{} on SOM and HOM success datasets}
  \label{tab:rq2hom}
  \begin{tabular}{lrrrr}
    \rowcolor{black!10}
    Task & SOM Simple & HOM Simple & SOM Complex & HOM Complex \\
    \noalign{\vskip 2pt}
        Localization - One Location & - & 100\% & - & 100\%\\
        Localization - First Location & - & 80\%  & - & 65.7\%\\
        Localization - Set Locations & - & 33.6\% & - & 32.8\%\\
        Localization - All Locations Strict &  82.6\% & 19.7\%  & 76.9\% & 25.3\% \\[5pt]
        
        Code Change Generation - Identical Code &   51.1\% & 2.4\% & 71.6\%  & 4.3\% \\
        Code Change Generation - Semantic Code Clone &  68.5\% & 49.2\% & 71.6\% & 55.4\%  \\[5pt]

        Behavior Change Prediction &  87.1\% & 96.8\% & 80.0\%  & 99.2\%   \\

\end{tabular}
\end{table}

Predicting \emph{behavior changes} has already been successfully achieved by the fine-tuned models on SOM. Nevertheless, the accuracy was still significantly increased on the HOM datasets. With more training data available, the models seem to better learn the effect of software changes. The higher complexity of changes does not hinder the model's understanding of the programs' behavior since up to 99.2\% accuracy was reached on the HOM\_Complex\_Success dataset.

Generating code to achieve the desired behavior has been the most challenging task overall. While for SOM, up to 71.6\% accuracy was achieved for semantic code clones, only up to 55.4\% of code changes were correctly generated for HOM. With the increasing change complexity, models struggle to achieve the desired behavior without side effects. In the used datasets, samples consisted of both code versions: one input and the corresponding output. Thus, for code change generation, in many cases, the generated code does produce the desired output but is not a semantic code clone of the expected code, as the test cases on one or a few other inputs don't match. Therefore, an additional experiment in \Cref{sec:rq6} will assess if providing a list with all the input-output pairs for a specific code sample will help to better represent the desired semantics. Nevertheless, being able to successfully predict the required code changes for up to 55.4\% of HOM and 71.6\% of SOM already significantly reduces manual investigations, especially because code changes are verifiable by running the changed code.

Changing code to achieve a desired behavior involves localizing where a change needs to be performed. While for SOM only a single location needs to be changed, for HOM several locations might need to be changed (in our setup, two or three locations). Changes in the program code might depend on each other; thus, in some cases, changes need to be performed in a strict order, as covered by the \emph{All Locations Strict} metric. For HOM, in 19.7\% (simple) and 25.3\% (complex) of the cases, all the locations that need to be changed were predicted in the correct order, as detailed in the localization part of \Cref{tab:rq2hom}. However, for most changes, the changes were required in distinct locations where modifications can be applied in an arbitrary order. Thus, the \emph{Set Locations} metric shows that for up to 33.6\% of the desired behavior changes, all the locations that need to be modified were successfully detected. Ultimately, it is desired that a model can successfully predict all the locations that need to be changed. However, correctly identifying where the first change needs to be applied already significantly reduces the search space. The fine-tuned models were able to accurately predict the first location in 65.7\% (complex) and  80\% (simple) for the HOM behavior changes. At least, a model should be able to target one location correctly that needs to be changed to achieve the desired behavior change which was the case for 100\% of the test samples. Thus, localizing at least one of the locations can be successfully achieved with enough training samples, while predicting all the locations can already be achieved for a third of the desired behavior changes.

\highlight{In this work, we highlight the potential of Neural Change Prediction to target higher order software changes allowing to localize at least one change location, generate the required code change for approximately half of the samples while showing excellent performance when predict the effect of software changes on a program's behavior.}

\subsection{RQ6: Behavior Context}\label{sec:rq6}

\emph{How does more context about the programs' dynamic behavior influence the models capabilities in learning to predict code changes?}
The behavior of a program can be expressed as the program's output for a given input. While this information is sufficient to learn the effect of software changes for a given input as seen in \Cref{sec:rq5}, it only provides limited context on the desired program behavior given a desired behavior change for a given input. Whether the program should behave the same or differently on other inputs is still unclear. Therefore, we further investigate if providing further context by adding several input-output pairs to express the previous and desired behavior, improves the learning of required code changes to achieve the desired behavior changes.

\subsubsection{Experimental Setup}
Expressing behavior changes can be done by specifying the desired output of a program, given a specific input, as done in the previous experiments. However, the semantics of a program can be too complex to be expressed in a single input-output pair. Thus, in this experiment, we evaluate how effectively a model can learn to generate the code changes for a desired behavior change expressed as a list of input-output pairs. The code samples for SOM and HOM remain the same as in the previously used ``success'' datasets. However, by combining the input-output pairs together for each code chunk, the total number of samples has reduced, especially for the simple projects where the programs were tested on up to 14 inputs. Thus, in this experiment SOM\_simple\_success\_pairs contains only 430 samples and HOM\_simple\_success\_pairs 29640 samples.
For the complex projects, the total amount of samples decreased as well even though the programs were only executed on two, three, or four inputs. Finally, the SOM\_complex\_success\_pairs resulted in only 10 samples, therefore no model was trained for SOM complex. For the complex HOM, 705 samples remained in HOM\_complex\_success\_pairs. For the evaluation, up to three generations were performed for code generation, same as in the previous experiment.

\subsubsection{Results}
In the previous experiment from \Cref{sec:rq5}, more correct code changes were generated for single order changes rather than for high order changes. Surprisingly, when providing more context about the program's behavior, for single order changes, the model performance decreased as can be seen in~\Cref{tab:rq6}. This is likely due to the limited amount of training samples. Due to the small sample size, the model is likely to focus only on a sub set of the input-output pairs from the list, leading to incorrect behavior. Additionally, LLMs have seen a huge amount of data with single input-output pairs and seem to be ``confused'' about this setup.

\begin{table}[h]
  \caption{Accuracy of fine-tuned \gpt{} with all input-output pairs}
  \label{tab:rq6}
  \begin{tabular}{lrrrr}
    \rowcolor{black!10}
    Task & SOM Simple & HOM Simple & SOM Complex & HOM Complex \\
    \noalign{\vskip 2pt}
        Code Change Generation - Identical Code &   25.6\% & 2.2\% & -  & 7.0\% \\
        Code Change Generation - Semantic Code Clone &  41.9\% & 53.5\% & - & 93.0\%  \\

\end{tabular}
\end{table}

Nevertheless, for HOM, the model's performance increased. For the projects with simple inputs, 53.5\% were achieved. Most interestingly, for the model fine-tuned on HOM\_complex\_success\_pairs, 93\% accuracy were reached. This highlights that adding additional behavior context can be beneficial in some cases. From the comparison between SOM simple and HOM simple, the performance improvement is likely due to the increase of available training samples which grew exponentially from 430 to 29.640 samples.
The most surprising performance difference remains between HOM simple and HOM complex. Here, the main difference is the amount of inputs in the test suite. While for the programs with simple input, up to 14 different input-output pairs were provided in the list, only up to four different input-output pairs were available for the programs with complex inputs. This might be an ideal amount of behavioral context to establish the general semantics of the program without inducing the model to focus to much on special corner cases that can cause the program to fail on more general inputs.

\highlight{More context about the programs' dynamic behavior can be beneficial if enough training samples to learn the relationship between code and behavior changes for several inputs are available. While this experiment shows some interesting insight into leveraging more behavioral context and its benefits for Neural Change Prediction, it also opens new path to explore further in future work.}

\subsection{Summary of Findings}
\Cref{sec:rq1} has shown the feasibility of Neural Change Prediction to learn and predict code changes based on desired behavior outcomes and vise versa. Further, \Cref{sec:rq2,sec:rq3,sec:rq4,sec:rq5,sec:rq6} investigated which factors impact the performance of the model.

\begin{itemize}
\item Overall, models that learned the effect of software changes are very confident to \emph{predict behavior changes} for SOM and HOM, achieving up to 99.2\% accuracy with no more than 1.04 attempts on average. Additionally, code changes that produce errors can be reliably detected and predicted with very little false positives (less than 1\%). 

\item Generating \emph{code changes} to obtain a specific behavior takes on average 1.56 attempts to get a semantic code clone for SOM with an accuracy up to 71.6\%. For code generation of HOM, the accuracy for generating programs consuming simple and complex inputs range around 50\%. However, when providing additional behavioral context, for HOM complex, 93\% of semantic code clones were correctly generated given a desired change in behavior. For both SOM and HOM code changes, programs consuming complex inputs are slightly easier to generate. Which is likely due to the fact that only a subset of the applied mutations lead to the successful execution of the program, which additionally came with a smaller test suite than the programs with simple inputs.

\item \emph{Localizing} where changes are required is the most challenging task for general LLMs as our baseline results showed in~\Cref{sec:rq1} by reaching only 9\% accuracy for localizing single change locations. Nevertheless, for this challenging task, we have seen the biggest improvement of leveraging Neural Change Prediction. For single locations, for programs with both simple and complex inputs, the locations were accurately predicted in 82.6\% and 76.9\% of the cases respectively. For HOM where several locations were changed, approaximatly 33\% of the cases, all locations that need to be changed to achieve a desired behavior change, were predicted. Nevertheless, for HOM where up to three locations have to be changed, for every sample, at least one location was correctly identified.
\end{itemize}
This work opens up new path to explore which will be further elaborated in~\Cref{sec:discussion}.

\highlight{The six experiments have highlighted the potential of Neural Change Prediction and its benefits for several software engineering tasks, ranging from feature and fault localization, over code generation to accurately predicting the effect of software changes on a programs' behavior.}

%% file: sections/discussion.tex
\section{Threats to Validity}\label{sec:discussion}
The current implementation of our fundamental and automated technique, Neural Change Prediction, represents a first study on its potential to learn the effect of software changes. 
Our experimental evaluation raised some threats to validity that will be discussed.

\subsection{External Validity}

\textbf{Benchmark.} For this study, we used the QuixBugs benchmark as a starting point for applying mutations.
While it is a commonly used benchmark for code generation and especially for bug fixing approaches, the benchmark itself is most probably known by LLMs. 
However, in this study, the risk for data contamination is small, as we used mutations to create our datasets.
This means that during the evaluation, we only evaluated the performance of the fine-tuned models in generating \emph{mutated versions} of the projects.
The same holds for predicting behavior changes, where we only predicted the behavior of the mutated projects. Although it is a commonly used benchmark for Python studies, it might not be representative enough of large scale software changes.

\textbf{Model architecture.} All experiments used OpenAI's \gpt{} model, while the model is currently available for fine-tuning, it cannot be guaranteed that the experiments can be reproduced in the future on the exact same model version which raises an external threat to validity. Nevertheless, the case study demonstrated that open source models can reach similar accuracy while learning the effects of software changes. Although the aim of the work was to provide a foundation for learning the effects of software changes, the model architecture was considered secondary. Other model architectures might produce slightly better results; however, the aim was not to investigate all available architectures but rather to demonstrate the general capabilities and new possibilities with Neural Change Prediction. 

\subsection{Internal Validity}
\textbf{Dataset construction.} To gather a large amount of training samples to learn the effects of software changes, all the mutation operators have been applied on every applicable node in the AST.
For single order mutants, all successfully executed samples have been included in the SOM\_Simple\_Success and SOM\_Complex\_Success.
However, for higher order mutants, we only kepta subset for training.
Due to the rapid mutant's explosion, we only kept mutants up to degree three were kept.
This means that the HOM datasets contain already mutants with complex changes that might have been applied in three distinct locations.
While equivalent mutants and mutants leading to failures have been omitted, the dataset still consisted of several hundred thousand samples.
As different subjects have varying numbers of lines of code, each subject lead to a different amount of mutants.
For degree three, those differences reached up to 100k mutants. Keeping all samples, would have significantly unbalanced the dataset.
For HOMs of degree two, all samples were kept since the total amount of samples was only approximately 36k samples for the simple subjects and of approx. 450 for the complex samples, with only minor imbalances between the subjects.
To balance the overall HOM dataset, for degree three, we set a maximum of 2500 mutants per subject.
This lead to the HOM\_Simple\_Success dataset containing 74,158 samples as well as the HOM\_Complex\_Success dataset with 1,201 samples.
While this process allowed gathering enough learning samples while keeping the dataset relatively balances, keeping a random subset of maximum 2,500 mutants per subject might not cover all the semantic changes leading to a maximized variety of behavior changes.

\subsection{Construct Validity}
\textbf{Evaluation metrics.} To evaluate the code generation task, inspiration was taken from the patch validation in automated program repair pipelines where a patch is validated by the test suite. Similarly, in our experiment, the generated code samples were executed on each input from the test suite. Only if all executions resulted in the same behavior as the expected code, they were considered semantic code clones. To evaluate fault localization, the literature typically measures top-1 and top-5 fault localization, meaning that the most suspicious line was either ranked first or at least in the top five.
For one generation attempt, checking if the predicted line matched the expected line corresponds to the top-1 localization.
For multiple attempts, we did receive five generation attempts, however, this is not strictly equal to the top-5 metrics as we did not request a ranking of the top five line and only got one prediction per generation attempt.
For effect prediction, the used metric was binary as we checked if the predicted behavior was equal to the expected behavior thus there is no threat to validity for that task.
In the case study, evaluation metrics were defined to assess the correctness of the generated CSS edits. Since, no established evaluation metrics for CSS were found in the literature, the Strict@ metric was used to validate exact syntactic matches while the Relaxed@ metric assessed the semantic similarity while validating that the desired behavior change occurred. Future work could especially improve on the IoU@ metric, established to calculate how accurately elements were swapped and thus located in the rendered website.

%% file: sections/rw.tex
\section{Related Work}\label{sec:rw}

Let us present relevant related work in the context of relating code changes and their effects, as well as automatically debugging varying software systems.

\subsection{Change Impact Analysis}
\emph{Debugging} software is about finding why the software does not produce the desired output and how to change it. In the literature, automated debugging is typically discussed in the context of debugging failures and investigating why a failure occurs. \emph{Change impact analysis}~\cite{arnold1996software} determines which parts of a program could be affected by a change. Being able to predict the effect of a code change has been known as an important task for decades as shown by Li et al.~\cite{li2013survey}. However, most works have focused on Java~\cite{ren2004chianti,ryder2001change} with only little research on other programming languages. With Neural Change Prediction we enable change impact analysis in a language agnostic way.

\subsection{Fault Localization}
Fault localization consists of identifying which parts of the program cause the failure. Several techniques have been developed to provide a ranking of the code locations that are the most likely to contain the bug. Spectrum-based fault localization (SBFL)~\cite{abreu2007accuracy} identifies suspicious code locations based on their occurrence in the execution traces of failing and passing test executions, different suspiciousness scores can be computed for the ranking such as e.g. the Ochiai score or Tarantula~\cite{jones2002visualization}. Once the likelihood of program entities to be faulty is assessed, this information can guide Automated Repair tools~\cite{liu2019you}.

\subsection{Automated Program Repair} Automated Program Repair (APR) aims at automatically changing software such that a given error no longer occurs. Traditionally, most APR tools operate on the buggy source code leveraging a correctness criteria such as the execution of a test suite. Once the potential buggy-ness of code locations has been identified, there are multiple APR techniques that can be applied.

\begin{description}
\item[Heuristic repair approaches]
    \cite{le2011genprog,liu2019tbar} rely on the amount of passing and failing tests as a fitness function to guide the repair process. Based on the suspiciousness scores, the most likely locations in the Abstract Syntax Tree (AST) are being mutated to create a patch candidate. Those are then evaluated against the test suite, based on how many tests are successfully executed, they are either further evolved or discarded. Once all the tests pass, the patch is considered valid. However, a patch that passes the test suite is not necessarily semantically equivalent to the developer intended fix and thus not always correct. To address the explosion of mutants that can be applied to the different suspicious locations, many APR tools generate patches based on fix templates~\cite{liu2019tbar,koyuncu2020fixminer}.
    Nevertheless, all those traditional techniques use a generate-and-validate approach~\cite{le2011genprog}, making the patch quality dependent on the existing test suite which leads to overfitting patches~\cite{yang2017better}.
    Due to the incompleteness of existing test suites, there is a need for additional specifications or dynamic execution data to improve the patch generation process.

    \item[Constraint-based techniques] \cite{nguyen2013semfix,mechtaev2016angelix} starts by performing symbolic execution to formulate a repair constraint that a patch candidate should satisfy. Later, an SMT solver is required to generate patches that satisfy the given constraint. Overall, constraint-based techniques still face scalability challenges while heuristic based approaches suffer from overfitting to the often incomplete test suites.

    \item[LLM-based APR approaches] \cite{zhang2023survey} recently have improved APR considerably. With the rise of LLMs, several studies~\cite{zhang2024systematic} assessed the capabilities of LLMs to fix bugs by providing the buggy code with varying artifacts such as a failing test case or a bug report as prompt. Additionally, retrieval-augmented LLM techniques like ReAPR~\cite{liu2025reapr} leverage bug fixes from historical data.

    However, simply querying an LLM only provides acceptable performance on simple well known cases and dramatically fails on real world bugs due to their lack of project specific and execution related context. ThinkRepair~\cite{yin2024thinkrepair} aims at addressing these limitations by applying Chain-of-Thoughts prompts with a collection and fixing phase. Since repair ingredients are essential for successfully fixing bugs, ReinFix~\cite{zhang2025repair} combines static analysis tools to assist the LLM and additionally leverages external ingredients collected from historical bug fixes with similar bug patterns.
    Nevertheless, the approach is not using dynamically retrieved execution data. RepairAgent~\cite{bouzenia2025repairagent} is an LLM agent capable of planning and executing tools to collect feedback on how to fix bugs. While agents can e.g. execute the program and run a test suite, they focus on repairing bugs by leveraging ingredients from previous fix attempts. However, they do not explicitly learn from the code changes impact on the behavior beyond bugs.
\end{description}

\subsection{Code Generation for and beyond APR}
Additionally, only little research has been done on investigating why a particular behavior occurs and how we can change it outside of failures. A recent study~\cite{shi2025natural} has investigated code generation throughout the software development process using natural language outlines of code. Although this allows to better represent the desired behavior of the code, the models do not learn how code changes affect the programs' behavior.

Generating code changes to achieve a particular behavior is a challenging task that does not only require static knowledge of the project but also dynamic information collected at runtime. Recent studies have highlighted the need for dynamic execution as well as repository level information. DynaFix~\cite{huang2025dynafix} integrates execution level dynamic information into their APR pipeline to fix Java programs from the Defects4J benchmark while RepoRepair~\cite{pan2026reporepair} highlights the need from repository level information for APR on SWE-bench Lite. 

Leveraging dynamic execution level and project specific information in the repair pipeline is a great first step. However, existing codebases are limited (even though huge) and have already been consumed by LLMs. While exiting LLMs demonstrate new possibilities to learn from code, it is essential to provide meaningful learning samples. Historically, mutations have been leveraged in mutation testing to assess the robustness of test suites~\cite{papadakis2019mutation} and used to perform targeted modification in the context of APR~\cite{le2016history} since small changes introduced in previous fixes can be abstracted in fix templates and applied through mutations. However, mutations can also be leveraged to gain a significant amount of diverse software projects which combined with their execution can be used to construct an unlimited amount of learning samples, targeted to the subject of interest. Now that we have reached the age of experience where AI agents gather project specific knowledge by injecting faults into a program~\cite{wei2025toward}, the need for learning from meaningful software changes is essential. While Self-play SWE-RL~\cite{wei2025toward} provides an interesting Agentic learning setup, code changes mostly consist of deletions that causes test cases to fail, while in our work, we leverage mutations to collect a huge amount of mutants which can be successfully executed and achieve a wide diversity of program behaviors that we can learn from, not only for APR but generally to generate code changes to achieve a specific behavior. Additionally, with Neural Change Prediction, it is possible to predict what the effect of specific code changes will be.

%% file: sections/conclusion.tex
\section{Conclusion and Future Work}
\label{sec:conclusion}

We introduce Neural Change Prediction, a novel approach to learn the association between code changes and their effects on program behavior, allowing to predict the effects of software changes, localize where to change code, and generate the necessary code changes.
To train its models, Neural Change Prediction applies large numbers of synthetic code changes and captures their effects on program behavior.
This training does not require knowledge of the underlying code language or its semantics; all it needs is a set of applicable mutations and observable (output) behavior.

While Neural Change Prediction is not bound to a particular model architecture or family, we demonstrate its feasibility by fine-tuning large language models on datasets of software changes; this allows us to benefit from the models' strong capabilities in understanding and generating code as well as interpreting natural language.
Our experiments show that Neural Change Prediction can achieve high accuracy in predicting the effects of software changes and predicting desired code changes, even when dealing with complex relationships between code and outputs---tasks that current LLM systems struggle with.
Our work thus represents a significant step towards enabling software developers (and AI systems!) to better understand and predict the effects of code changes, as well as to suggest the necessary code changes to achieve desired effects, ultimately improving software development and maintenance processes.

While our results highlight the potential of Neural Change Prediction, they also open up many paths for future work:
\begin{description}
    \item[Larger subjects.] We plan to apply Neural Change Prediction to \emph{larger codebases} and \emph{multi-language open-source projects,} to further validate its effectiveness and scalability.
    \item[More complex changes.] We will explore the application of Neural Change Prediction to \emph{more complex code changes,} such as refactorings and features added, and investigate how well it can learn from and predict such changes.
    \item[Behavior representation.] Expressing the behavior of a program in a format that can be leveraged by the model and a human is a challenging task. We will explore how to represent \emph{more detailed runtime behavior} with execution features to express more fine-grained behavioral changes. 
    \item[Model architectures.] We will investigate the use of \emph{alternate model architectures,} such as graph neural networks or local transformer-based models, to better capture the relationships between code changes and their effects on behavior, and to reduce the reliance on large language models.
\end{description}

To encourage further research in this area and to allow others to build upon our work, all our code, datasets, and experimental results are available on request.